\documentclass[sigconf,fontsize=9pt,nonacm]{acmart}

\AtBeginDocument{%
}

\usepackage{colortbl}

\usepackage{graphicx}
\usepackage{grffile}
\usepackage{tikz}
\usetikzlibrary{calc}

\usepackage[english]{babel}
\usepackage[linguistics]{forest}

\usepackage{fontawesome}

\usepackage{booktabs}
\usepackage{tabularx}
\usepackage{multirow,multicol,threeparttable, tablefootnote}

\usepackage{subcaption}
\usepackage{comment}

\usepackage{hyperref}
\usepackage{mathtools}
\usepackage{bm}

\usepackage{empheq}

\usepackage{enumitem}

\usepackage[export]{adjustbox}

\usepackage{lineno}

\usepackage{amsthm}

\usepackage{pgf}

\newdimen\mydim

\newcommand{\gettikzxy}[3]{%
  \tikz@scan@one@point\pgfutil@firstofone#1\relax
  \edef#2{\the\pgf@x}%
  \edef#3{\the\pgf@y}%
}

\newdimen\XCoord
\newdimen\YCoord

\DeclarePairedDelimiter\ceil{\lceil}{\rceil}
\DeclarePairedDelimiter\floor{\lfloor}{\rfloor}

\usepackage[ruled,vlined,linesnumbered,resetcount,algosection,nofillcomment]{algorithm2e}

\SetKwInput{KwData}{Input}
\SetKwInput{KwResult}{Output}
\SetKwFor{Pfor}{parallel for}{}{}
\SetKwFor{PipFor}{pipelined for}{}{}
\SetKwFor{PipForEach}{pipelined foreach}{do}{endfch}
\SetKw{KwBy}{by}
\SetKw{KwNot}{not}
\SetKw{KwBreak}{break}
\SetKwFor{Loop}{loop}{}{}

\SetKwRepeat{Do}{do}{while}
\SetKwRepeat{PipDo}{pipelined do}{while}
\SetKwProg{Pn}{function}{:}{\KwRet}
\SetKw{Output}{output}

\let\oldnl\nl
\newcommand{\nonl}{\renewcommand{\nl}{\let\nl\oldnl}}

\SetCommentSty{mycommfont}

\usepackage{cleveref}

\usepackage{resizegather}

\usepackage{soul}

\usepackage[skip=2pt]{caption} 
\usepackage[short]{optidef}

\usepackage{xcolor}

\colorlet{pink}{red!40}
\colorlet{cyanblue}{cyan!80}

\newcommand*{\tikzmk}[1]{\tikz[remember picture,overlay,blend mode=multiply] \node (#1) {};\ignorespaces}

\newcommand{\boxfor}[1]{\tikz[remember picture,overlay,blend mode=multiply]{\path let \p1 = (Zero) in node [yshift=3pt,xshift=6pt,fill=#1,opacity=.40,fit={(Start)($(.92\linewidth + \x1,-.5\baselineskip) $)}] {};}\ignorespaces }

\newcommand{\boxline}[1]{\tikz[remember picture,overlay,blend mode=multiply]{\path let \p1 = (Zero) in node [yshift=3pt,xshift=6pt,fill=#1,opacity=.40,fit={(Start)($(.92\linewidth + \x1,.8\baselineskip) $)}] {};}\ignorespaces }

\definecolor{myred}{RGB}{250,214,221}
\definecolor{myblue}{RGB}{190,228,254}

\newcommand{\unsim}{\mathord{\sim}}

\usepackage{pifont}

\pgfmathsetmacro{\nodebasesize}{1} 
\pgfmathsetmacro{\nodeinnersep}{0.05}

\newcommand*\circled[1]{\tikz[baseline=(char.base)]{
        \node[shape=circle,draw,minimum size=1.5mm, inner sep=0pt,line width=0.25mm] (char)
        {\rule[-1.8pt]{0pt}{\dimexpr1.5ex+1.5pt}\small \textbf{#1}};}}

\newcommand\rv[1]{{#1}}

\newcommand{\setword}[2]{%
  \phantomsection
  #1\def\@currentlabel{\unexpanded{#1}}\label{#2}%
}

\newcommand{\mref}[2]{\textcolor{red}{\hyperref[#1]{#2}}}

\def\@copyrightspace{\relax} 
\setcopyright{none}
\renewcommand\footnotetextcopyrightpermission[1]{} 
\begin{document}

\title{LightRW: FPGA Accelerated Graph Dynamic Random Walks}

\author{Hongshi Tan}
\email{hongshi@comp.nus.edu.sg}
\orcid{0000-0002-3243-6875}
\affiliation{%
  \institution{National University of Singapore}
  \country{Singapore}
}

\author{Xinyu Chen}
\email{xinyuc@comp.nus.edu.sg}
\orcid{0000-0003-1951-5015}
\affiliation{%
  \institution{National University of Singapore}
  \country{Singapore}
}

\author{Yao Chen}
\email{yaochen@nus.edu.sg}
\orcid{0000-0002-5798-2282}
\affiliation{%
  \institution{National University of Singapore}
  \country{Singapore}
}

\author{Bingsheng He}
\email{hebs@comp.nus.edu.sg}
\orcid{0000-0001-8618-4581}
\affiliation{%
  \institution{National University of Singapore}
  \country{Singapore}
}

\author{Weng-Fai Wong}
\email{wongwf@nus.edu.sg}
\orcid{0000-0002-4281-2053}
\affiliation{%
  \institution{National University of Singapore}
  \country{Singapore}
}

\renewcommand{\shortauthors}{Hongshi Tan et al.}

\begin{abstract}
Graph dynamic random walks (GDRWs) have recently emerged as a powerful paradigm for graph analytics and learning applications, including graph embedding and graph neural networks.
Despite the fact that many existing studies optimize the performance of GDRWs on multi-core CPUs, massive random memory accesses and costly synchronizations cause severe resource underutilization, and the processing of GDRWs is usually the key performance bottleneck in many graph applications. This paper studies an alternative architecture, FPGA, to address these issues in GDRWs, as FPGA has the ability of hardware customization so that we are able to explore fine-grained pipeline execution and specialized memory access optimizations.
Specifically, we propose {LightRW}, a novel FPGA-based accelerator for GDRWs. LightRW embraces a series of optimizations to enable fine-grained pipeline execution on the chip and to exploit the massive parallelism of FPGA while significantly reducing memory accesses. As current commonly used sampling methods in GDRWs do not efficiently support fine-grained pipeline execution, we develop a parallelized reservoir sampling method to sample multiple vertices per cycle for efficient pipeline execution. To address the random memory access issues, we propose a degree-aware configurable caching method that buffers hot vertices on-chip to alleviate random memory accesses and a dynamic burst access engine that efficiently retrieves neighbors. Experimental results show that our optimization techniques are able to improve the performance of GDRWs on FPGA significantly. Moreover, LightRW delivers up to $9.55{\times}$ and $9.10{\times}$ speedup over the state-of-the-art CPU-based MetaPath and Node2vec random walks, respectively. \rv{This work is open-sourced on GitHub at \url{https://github.com/Xtra-Computing/LightRW}.}

\end{abstract}

\ccsdesc[500]{Hardware~Hardware accelerators}
\ccsdesc[500]{Information systems~Query operators}
\ccsdesc[300]{Computer systems organization~Data flow architectures}
\ccsdesc[300]{Mathematics of computing~Graph algorithms}

\keywords{FPGA accelerator; parallel weighted reservoir sampling; random walk on graphs.}

\maketitle

\section{Introduction}
\label{sec:introduction}

\begin{figure}[t]
\centering
\centering
\includegraphics[width=\linewidth]{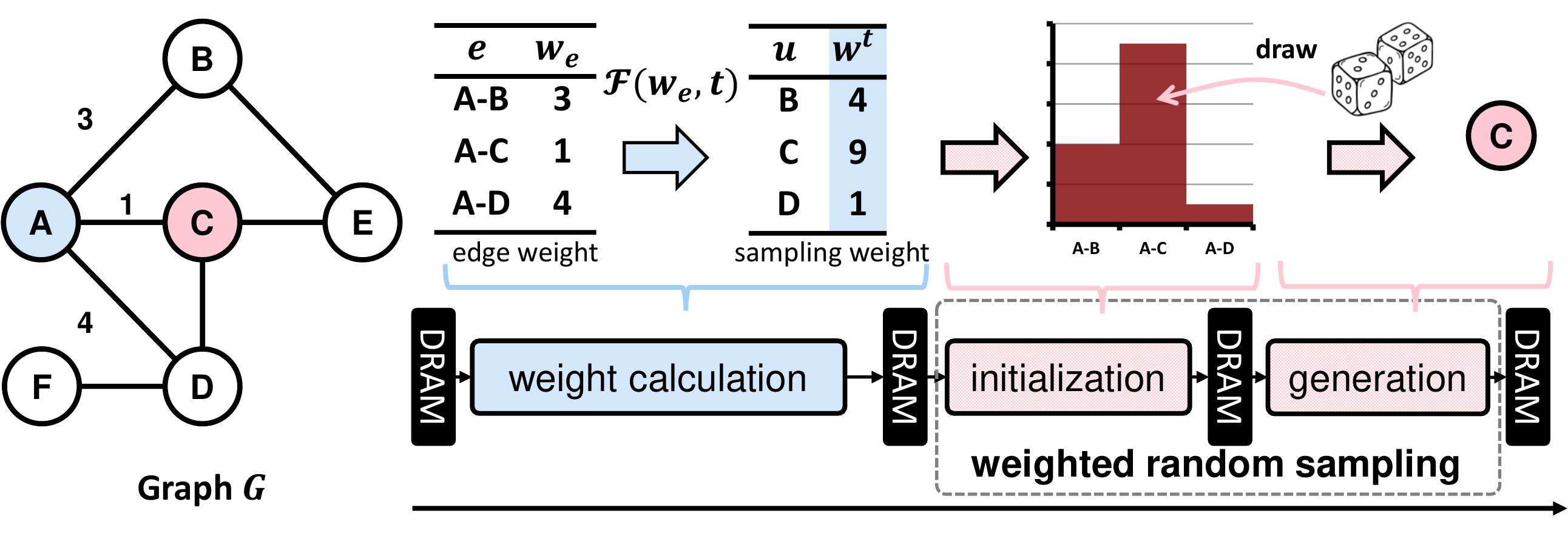}

\caption{Processing flow of GDRW on CPUs.}
\label{fig:processflow_cpu}
\vspace{-2mm}
\end{figure}

\rv{\setword{Random}{rv:r1o2_1} walk on graphs has been widely adopted by many applications, such as the recommendation system~\cite{gori2006research}, bioinformatics~\cite{wang2021essential}, and network analysis~\cite{lu2011link}. Due to the ineffectiveness of traditional random walk algorithms (i.e., uniform random walk and DeepWalk~\cite{perozzi2014deepwalk}, etc.) in extracting temporal relations with structural information of the graph, graph dynamic random walks (GDRWs) have attracted increasing attention~\cite{grover2016node2vec,dong2017metapath2vec,wan2020reinforcement,he2019hetespaceywalk,vahedian2016weighted,nikolentzos2020random}. For example, Node2Vec~\cite{grover2016node2vec} has been widely used to generate graph embeddings for downstream machine learning models~\cite{nikolentzos2020random}, and MetaPath~\cite{dong2017metapath2vec} plays an important role in knowledge graph analytics~\cite{yu2014personalized}.
}
However, due to inherent random memory accesses and high computational complexity, GDRWs generally dominate the overall execution time of these applications.
For example, the state-of-the-art approach, ThunderRW by Sun et al.~\cite{sun2021thunderrw}, showed that Node2Vec~\cite{grover2016node2vec} can take up to two hours on a graph of 41 million vertices and 1.21 billion edges on a modern multi-core machine.
Due to the explosive growth of graph data and the rising popularity of GDRWs in numerous applications, there is an urgent need to accelerate the performance of GDRWs.

Extensive efforts~\cite{sun2021thunderrw,yang2019knightking} have been made to optimize GDRWs on multicore CPUs. For example,
\citeauthor{yang2019knightking}~\cite{yang2019knightking} explore the parallelism of GDRWs in a distributed environment by efficient graph partitioning techniques.
ThunderRW~\cite{sun2021thunderrw} develops a stage design and interleaves memory accesses from massive queries for high memory efficiency and achieves state-of-the-art performance on modern CPU hardware.
\rv{Random walk queries are processed concurrently with multiple threads, and each thread handles different queries with the execution flow illustrated in Figure~\ref{fig:processflow_cpu}. For a query starting from vertex \emph{A} in graph \emph{G} to move one step forward, \textit{weight\_calculation} traverses all neighbors of \emph{A} at the start and then calculates the sampling weights using an application-specific weight update function $\mathcal{F}$, e.g., the sampling weights of vertex \emph{B}, \emph{C} and \emph{D} (\{4,9,1\}) are calculated by $\mathcal{F}$ with the corresponding edge weights \{3,1,4\} and the current state $t$. Then, weighted random sampling is conducted in two procedures. First, \textit{initialization} builds an indexing table from the sampling weight, which describes the distribution of all neighbors and is stored in global memory. Second, \textit{generation} generates uniformly distributed random samples to index the corresponding vertex. Then, the indexed vertex \emph{C} is marked as the starting vertex for the next step.
}
This process is repeated until it satisfies a specific termination condition, such as a target length being reached.

Despite those existing studies and optimizations for multi-core architectures, there are fundamental limitations in microarchitecture level to implementing GDRWs on multicore CPUs efficiently.
Our comprehensive profiling on ThunderRW shows that even with advanced optimizations such as explicit memory prefetching, $59.9\%$ of execution cycles are still stalled by memory accesses, with a cache miss ratio as high as $58.2\%$ (see more details in~\Cref{sub:inefficiency}). The irregular nature of graphs~\cite{safro2009improving} results in GDRWs having extremely poor spatial and temporal locality, rendering the multicore's complex cache hierarchy ineffective. Moreover, GDRWs on multicore architectures have to be executed in stages, with barriers required at the end of each stage. 
Intermediate data need to be written to DRAM after the execution of one stage and read again for the next stage. %

This paper studies an alternative architecture, FPGA, to address those issues in GDRWs. Different from the multicore CPUs that have fixed architectures with a deep cache hierarchy, FPGAs have the flexibility to customize hardware logic, which allows us to explore novel execution schemes such as pipeline execution on chip with fine-grained inter-stage communication (referred to herein as fine-grained pipeline execution) to minimize data movement overhead and design specialized memory engines to handle random memory accesses in GDRWs efficiently. Specifically, we propose {LightRW}, the first FPGA-based acceleration framework for efficient GDRWs.
LightRW creates a fully pipelined and parallelized FPGA-based accelerator system for GDRWs.
The major contributions of our LightRW framework are as follows:
\begin{itemize}[leftmargin=*]
\setlength\itemsep{-0.5mm}
    \item We design a parallelized Weighted Reservoir Sampling (WRS) on FPGA, which is able to process multiple vertices per cycle and enables fine-grained communication between stages so that the different stages are executed in parallel and communicate within the chip without retrieving the DRAM.
    \item We propose two memory efficient techniques to handle notorious random accesses to neighbors of vertices: 1) a degree-aware cache keeps high-degree vertices associated data (e.g., address of their neighbors) into on-chip memory at runtime, and 2) a burst access engine adjusts the burst access size to minimize redundant data access for higher memory bandwidth utilization. 
    \item Our evaluation of LightRW on a server with an FPGA board attached to the PCIe interface outperforms the state-of-the-art CPU-based implementation~\cite{sun2021thunderrw} by up to $9.55{\times}$ and $9.10{\times}$ on MetaPath~\cite{sun2013mining} and Node2Vec~\cite{grover2016node2vec}, even with the consideration of the data transfer overhead over PCIe, respectively. Furthermore, LightRW is $15{\times}\sim26{\times}$ more power efficient than the state-of-the-art CPU-based implementation.
\end{itemize}

The rest of the paper is organized as follows. We introduce the background and motivation in~\Cref{sec:background}.
~\Cref{sec:lightrw} presents the overview of LightRW, including the proposed pipelined GDRW algorithm and hardware architecture.
While ~\Cref{sec:parallelWRS} illustrates the design of the parallel WRS sampler, \Cref{sec:memory_optimizations} introduces two GDRW specific memory optimizations.
We present the experimental results in ~\Cref{sec:evaluation} and related works in~\Cref{sec:related_works}.
This paper is concluded in~\Cref{sec:conclusion}.

\setlength{\textfloatsep}{4pt}
\setlength{\dbltextfloatsep}{4pt}

\section{Background and Motivation}\label{sec:background}

In this section, we introduce the background of graph dynamic random walks, discuss the state-of-the-art CPU-based solutions and motivate our FPGA-based accelerator design according to the analysis on performance bottlenecks of CPU-based solutions.

\subsection{Graph Dynamic Random Walks}
Let $G=\{{V},{E}\}$ represent a directed graph with $|{V}|$ vertices and $|{E}|$ edges. \rv{$N({u})$ is refered to a set of all neighbors of ${u}$, and $|N({u})|$ be the total number of neighbors (the \setword{degree}{rv:r1o3}). Let $d_{{u}}^+$ and $d_{{u}}^-$ be the out-degree and in-degree of ${u}$ respectively} The undirected graphs are supported by representing each undirected edge with two directed edges with the same two vertices.
The edge between the vertex ${u}$ and the vertex ${v}$ is represented as $({u}, {v})$, and the corresponding edge weight is represented by $w_{{u},{v}}^*$, in the case of unweighted graphs, $w_{{u},{v}}^* =1$.

A random walk on a graph begins at a starting vertex and moves to a randomly selected neighbor of the currently residing vertex at each step. Based on how the edge (or the neighbor) is selected, random walks are further categorized into \textit{unbiased} or \textit{biased} ones. The transition probability of an edge is defined as the probability of the corresponding edge being sampled, and we refer to the unnormalized form of the transition probability as the sampling weight. For unbiased random walks, the transition probability is uniformly distributed, while for biased random walks, the transition probability is generally related to the edge weight. Furthermore, for biased random walks, if the edge weights remain constant throughout the process, they are called \textit{static} random walks.
In contrast, \textit{graph dynamic random walks} (GDRWs) take into account the walker's state by updating the edge weights at each step to recalibrate the transition probability. As in static random walks, per-edge transition probabilities could be calculated offline, and many existing studies take advantage of this to simplify online calculation~\cite{sun2021thunderrw, yang2019knightking, perozzi2014deepwalk}. In comparison, GDRWs require more advanced and challenging optimizations at runtime, which are the focus of this paper. Moreover, with the modeling on the walker's state, GDRWs are able to extract higher-order structural information and temporal relationships in graphs. Therefore, GDRWs have attracted increasing attention for graph learning applications.

Next, we introduce two representative GDRW algorithms: MetaPath and Node2Vec. \rv{Let $w_{{a},{b}}^t$ be the sampling weight of edge $(a,b)$ to be sampled from the current vertex $a$ at step $t$.} $w_{{a},{b}}^t$ is updated by an application-specific weight update function $\mathcal{F}$, i.e., $w_{{a},{b}}^t = \mathcal{F}(w_{{a},{b}}^*,V_{t-1})$, where $V_{t-1}$ is the set of traversed vertices.

\noindent\textbf{MetaPath}~\cite{sun2013mining} is an effective approach for data mining on heterogeneous graphs.
A MetaPath $M$ is defined as $L_1 \xrightarrow[]{R_1} L_2 \xrightarrow[]{R_2} ...\xrightarrow[]{R_t} L_{t+1}$, where $L_i$ is the vertex label in the $i$-th step, and $R_i$ is the relation between $L_i$ and $L_{i+1}$ (e.g., a book is cited by an author).
A MetaPath random walk query returns a sampled path based on the given relation path $\mathcal{R}={R_1,R_2,...,R_t}$. The weight update function is shown in~\Cref{eq:metapath}. \setword{}{rv:r1o2_3}
For the $t$-th step, 
For the $t$-th step, \rv{the weight of the path to be sampled is set according to whether the relationship is met, as shown in~\Cref{case:mt1}. Otherwise, the corresponding weight is set to zero, indicating that the path will not be sampled in this step (\Cref{case:mt2}).}
\begin{subequations}
    \begin{empheq} [left={w_{{a},{b}}^t=}\empheqlbrace] {align}
        w_{{a},{b}}^*, & ~~~\text{if ($({a},{b}) \in E )\land (R_{({a},{b})} = \mathcal{R}[t])$}, \label{case:mt1}\\
        0, & ~~~\text{otherwise}.\label{case:mt2}
    \end{empheq}
    \label{eq:metapath}
\end{subequations}
\noindent\textbf{Node2Vec}~\cite{grover2016node2vec} is a second-order random walk~\cite{ching2013higher} and has been widely used as a graph embedding technique~\cite{grohe2020word2vec}.
The transition probability depends not only on the current vertex ${a}^{t}$ but also on the previously traversed vertex ${a}^{t-1}$.
The weight of neighbors depends on whether it is connected to the last traversed vertex.
\rv{In cases where the neighbor $b$ is the same as the vertex visited in the previous step, the sampling weight is equal to the scaled edge weight with a hyperparameter $p$ (\Cref{case:nv1}). Apart from that, if there is an edge between ${a}^{t-1}$ and $b$, the sampling weight is equal to the edge weight (\Cref{case:nv2}); otherwise, the sampling weight is scaled with another factor $q$ in \Cref{case:nv3}. Finally, for all non-neighboring vertices, the sampling weights are set to zero (\Cref{case:nv4}).}
\begin{subequations}
    \begin{empheq} [left={w_{{a},{b}}^t=}\empheqlbrace] {align}
        \frac{w_{{a},{b}}^*}{p} , &~~~\text{if ${b} ={a}^{t-1}$}, \label{case:nv1} \\
        w_{{a},{b}}^* , & ~~~\text{if $(  {b}\neq {a}^{t-1})\land(({a}^{t-1},{b}) \in E),$} \label{case:nv2}~~~\\
        \frac{w_{{a},{b}}^*}{q} , &~~~\text{if $(  {b}\neq {a}^{t-1})\land(({a}^{t-1},{b})     \notin E),$} \label{case:nv3}\\
        0 , & ~~~\text{otherwise} \label{case:nv4}.
    \end{empheq}
    \label{eq:node2vec}
\end{subequations}

\subsection{GDRW on CPUs}

\begin{algorithm}[t]
\SetInd{0.1em}{2em}
\KwData{$G$: a given graph,
$\mathbb{Q}$: a set of input queries.
}
\KwResult{$V_{t}$: a path of traversed vertices.}
\tikzmk{Zero}
$\textit{res} = \emptyset$\;
\ForEach{$Q\in \mathbb{Q}$}{
    $v_{\textit{curr}}= Q.v_{\textit{start}}$,
    $Q.\textit{res} = \emptyset$\;
\Loop{}{

    $w_{\textit{sum}} = 0$,
    $W = \emptyset$,
    $N_v = \emptyset$ \;
    \tcc{weight\_calculation}
    \tikzmk{Start}$\{\textit{address}, \textit{degree}\} = \text{\textbf{get\_neighbors\_info}}(v_\textit{curr},G)$\;\label{line:1_load_ni}
    \For{$i\gets1$ \KwTo $\textit{degree}$}{
        $N_v^i = \text{\textbf{get\_neighbor}}(\textit{address}, i, G)$\;\label{line:1_load_prop}
        $w_v^i = \text{\textbf{app\_weight\_update}}(N_v^i, v_\textit{curr})$\;\label{line:1_weight}
        \tikzmk{End}\boxfor{myblue}
        $W.\textbf{push}(w_v^i)$\;
    }
    \tcc{weighted\_sampling}
    \tikzmk{Start}{$T = \text{\textbf{initialization}}(W)$ \tcp*{O(n) time complexity} \label{line:1_sample_init}
    $v_\textit{curr} = \text{\textbf{generation}}(T)$\;\label{line:1_sample}
    }\tikzmk{End}\boxline{myred}
    $Q.\textit{res}.\textbf{push}(v_\textit{curr})$\; \label{line:1_update}
    \If{$(Q.{\normalfont \textbf{is\_end}}())$}
    {
        $\textit{res}.\textbf{push}(Q.\textit{res})$\;
        \KwBreak \;
    }
}
}
\Return{$\textit{res}$}\;
\caption{CPU-based GDRWs}\label{alg:drw_nrs}
\end{algorithm}

The execution flow of the CPU-based GDRW~\cite{sun2021thunderrw} is shown in~\Cref{alg:drw_nrs}.
For an input query composed of a starting vertex $v_\textit{curr}$ and the requested length, the algorithm first ascertains the number of neighbors of the current vertex $v_\textit{curr}$ for indexing in~\Cref{line:1_load_ni}.
For each neighbor of $v_\textit{curr}$, the algorithm first loads its properties (\Cref{line:1_load_prop}) and then updates the weight of each neighbor using the application-specific function (\Cref{line:1_weight}). Different GDRW applications can be implemented by customizing the application-specific weight update function.
The updated weights are used to implicitly calculate the transition probability, which is the probability of the neighbors to be selected.

In the next stage, weighted sampling draws one sample based on the generated weight distribution $W$.
This process is carried out in two phases: \emph{initialization} and \emph{generation}, which are commonly used in sampling methods including inverse transformation sampling~\cite{sun2021thunderrw} and alias sampling~\cite{hubschle2019parallel}.
Since the updated weight is discretely distributed, the initialization stage constructs an intermediate table $T$ that describes the distribution of $W$, which allows for the sampling of weighted items with uniformly distributed random numbers (\Cref{line:1_sample_init}).
Finally, in~\Cref{line:1_sample,line:1_update}, the generation stage randomly draws one neighbor based on $T$ and appends it to the output sequence $Q.res$, which is then set as $v_\textit{curr}$ for the next iteration. This process is repeated until $v_{curr}$ satisfies the given terminal state (e.g., the required path length).
The state-of-the-art CPU-based system~\cite{sun2021thunderrw} adopts multiple threads to process different batches of queries in parallel. Each stage of a query is executed sequentially and interleaved with memory accesses using one thread, such that the latency of random access overlaps with computation.

\subsection{Inefficiencies of CPU-based GDRWs}\label{sub:inefficiency}

To further investigate performance potentials on multi-core architectures, we have conducted a top-down performance analysis~\cite{vtune} of the state-of-the-art CPU-based system, ThunderRW~\cite{sun2021thunderrw}. 
In particular, we evaluated MetaPath and Node2Vec with two widely used graphs in previous studies, \rv{livejournal~\cite{snapnets} and uk-2002~\cite{BoVWFI}}, on a server equipped with the latest Intel Xeon Gold 6246R CPU.
We used vTune~\cite{vtune} to profile the last-level cache miss ratio (denoted as "LLC Miss"), the proportion of cycles stalled by memory accesses (denoted as "Memory Bound"), and the proportion of cycles used for useful computations (denoted as "Retiring Ratio"). More details about the experimental setup can be found in~\Cref{sec:evaluation}.

The profiling results show that the LLC miss ratio is very high (up to $76.9\%$), and the memory bound ranges from $31.2\%$ to $59.9\%$ for the two graphs in both applications. Subsequently, the retiring ratio is only $8.2\% \sim 33.6\%$, indicating poor utilization of CPU cores.
In summary, memory accesses dominate the overall execution time for CPU-based GDRWs.
Based on the execution flow on CPUs and the characteristics of GDRW, there are two key observations for the inefficiency:

\begin{table}[t]
\centering
\caption{Profiling results of the state-of-the-art CPU-based system~\cite{sun2021thunderrw} on MetaPath and Node2Vec.}
\label{tab:cpu_benchmark}

\begin{tabular}{lclcc}
\toprule
\multicolumn{1}{c}{Applications}&
\multicolumn{1}{c}{Graphs}  & \begin{tabular}[c]{@{}c@{}}LLC\\ Miss\end{tabular}  & \begin{tabular}[c]{@{}c@{}}Memory\\ Bound\end{tabular} & \begin{tabular}[c]{@{}c@{}}Retiring\\ Ratio\end{tabular} \\

\midrule

\multicolumn{1}{l}{\multirow{2}{*}{MetaPath}}&
\multicolumn{1}{l}{liveJournal~\cite{snapnets}}  
&$\bm{58.2\%}$  &$\bm{59.9\%}$
&$\bm{8.2\%}$ \\
\cmidrule(lr){2-5}

\multicolumn{1}{l}{}&
\multicolumn{1}{l}{uk-2002~\cite{BoVWFI}} 
&$\bm{61.8\%}$  &$\bm{57.5\%}$
&$\bm{13.7\%}$ \\
\midrule

\multicolumn{1}{l}{\multirow{2}{*}{Node2Vec}}&
\multicolumn{1}{l}{liveJournal~\cite{snapnets}}  
&$\bm{76.9\%}$  &$\bm{31.2\%}$
&$\bm{23.3\%}$ \\
\cmidrule(lr){2-5}

\multicolumn{1}{l}{}&
\multicolumn{1}{l}{uk-2002~\cite{BoVWFI}} 
&$\bm{61.1\%}$  &$\bm{31.7\%}$
&$\bm{33.6\%}$ \\

\bottomrule
\end{tabular}%

\end{table}

\noindent\textbf{Inefficiency 1: The sequential execution flow on CPUs introduces a large number of memory accesses.} The three stages of GDRW are executed in sequence with barriers required at each stage on control flow-based multicore CPUs.
Here, we provide a quantitative analysis of the number of memory accesses introduced by intermediate data access. The weight update stage of the neighbors requires storing updated weights with the size of $|N(v_\textit{curr})|$ in memory.
During the initialization stage, the updated weights are read and, a temporary data structure $T$ is created with a size of $|N(v_\textit{curr})|$.
Therefore, in addition to the output results, the entire process ideally requires $2{\times}|N(v_\textit{curr})|$ memory accesses. However, since the computations involved are typically lightweight, the costs of memory access cannot be overlapped. As a result, memory access tends to dominate the overall execution time.

\noindent\textbf{Inefficiency 2: Irregular memory accesses are poorly handled.}
Due to the irregular nature of graphs, GDRW introduces massive irregular memory accesses, including random access to the addresses of neighbors of a vertex and access to varying numbers of neighbors, leading to poor data locality.
\rv{\setword{Additionally}{rv:r1o2_4}, the size of intermediate data for existing weighted sampling methods (i.e., alias table~\cite{hubschle2019parallel} and inverse transform distribution table~\cite{enwiki:1115190568}, etc.) is proportional to the number of neighbors. For large graphs, the space required to cache neighbors of different nodes varies significantly, which easily exceeds the capacity of any CPU cache, leading to cache thrashing.
This results in significant cache misses and CPU core stalls.
Moreover, with multiple queries executed with multiple threads, concurrent prefetching in the shared last-level cache exacerbates the memory access contention and cache thrashing.}

\subsection{Motivation and Design Rationales}
\label{sub:potentials_of_fpga_based_accelerators}
The inefficiencies mentioned above are fundamentally caused by the sequential execution flow on multicore architectures and the mismatch between the memory access pattern of GDRW and the current cache hierarchy of multi-core architectures. It is clear that a "one size fits all" approach is not suitable. There is a need for a specialized solution in terms of both the algorithm and architecture. This paper studies radical and alternative architectures, starting with FPGA, a commodity hardware for specialized designs.

With the flexibility to customize hardware logic, FPGAs have demonstrated promising performance in many data-intensive applications~\cite{hu2021graphlily, chen2019fly,chen2021skew,chen2021thundergp,chen2022thundergpr,zhou2019hitgraph}. 
Among them, a common optimization technique is fine-grained pipelining, which instantiates hardware units for different tasks (e.g., functional stages in GDRW) on the chip and connects hardware units through FIFOs for fine-grained communication (e.g., a vertex per transfer). As a result, it minimizes the number of DRAM memory accesses and increases parallelism by simultaneous execution of hardware units in comparison to non-pipelined systems.
Meanwhile, customization of the memory access engine (i.e., using scratchpad instead of a complex cache hierarchy on CPUs) is another source of performance improvement. For example, the FPGA-based graph processing framework from Chen et al.~\cite{chen2021thundergp} outperforms state-of-the-art solutions by pipelining the Scatter and Gather functions of the GAS model~\cite{chen2021thundergp} and adopting application-specific memory access units. 
This motivates us to explore two directions to address the inefficiencies encountered in CPU-based GDRWs:
\begin{itemize}[leftmargin=*]
    \item Pipelining different stages of GDRW on FPGAs with fine-grained inter-stage communication to eliminate synchronization barriers and minimize data movement to DRAM.
    \item Designing specialized memory access engines with considerations on the irregular memory access patterns in GDRWs for high memory bandwidth utilization.
\end{itemize}

\section{LightRW}
\label{sec:lightrw}
With the motivation and design rationales, we introduce LightRW, an FPGA-based accelerator for efficient GDRW.

\subsection{Solution Overview}

\begin{figure}[t]
\centering
\includegraphics[width=\linewidth]{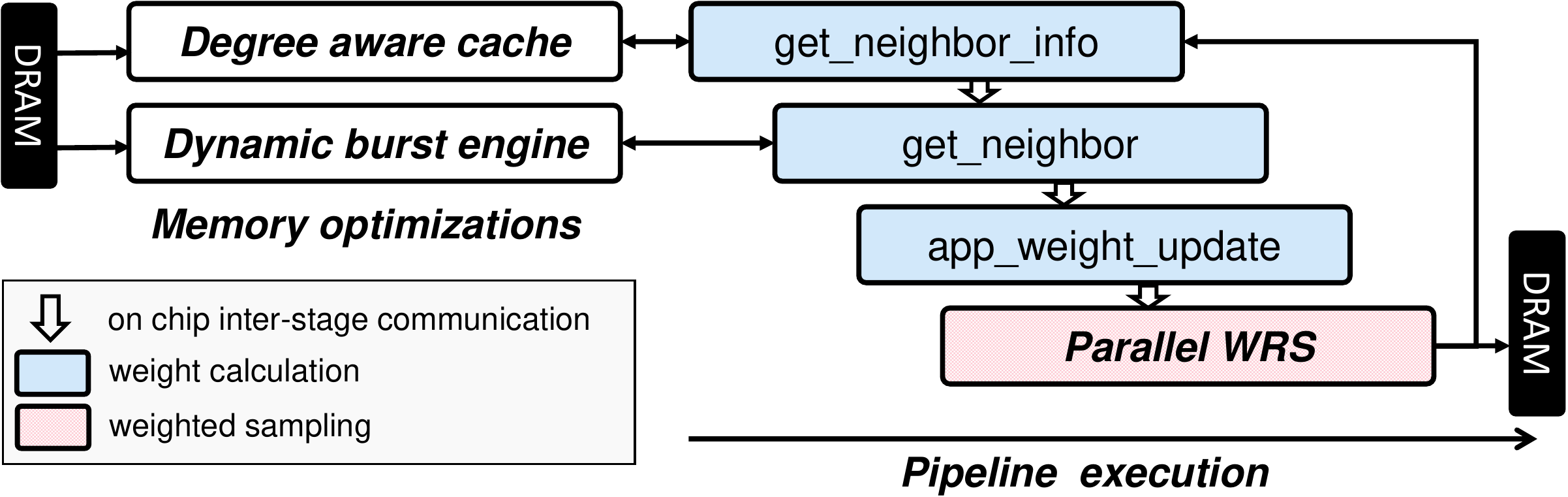}
\vspace{2pt}
\caption{Processing flow of LightRW.}
\label{fig:overviewlightrw}
\end{figure}

In short, the design of LightRW has two complementary approaches: 1) reducing the number of memory accesses to DRAM by enabling fine-grained pipeline execution of GDRW on FPGAs, and 2) handling random memory accesses efficiently with customized memory optimizations. Its processing flow is shown in ~\Cref{fig:overviewlightrw}.

LightRW eliminates the synchronization barriers of existing GDRW algorithms to enable fine-grained pipeline execution on the chip by adopting the {\em weighted reservoir sampling} (WRS) technique that chooses a random sample in a single pass over the items.
More importantly, we parallelize WRS to process multiple vertices per cycle for high throughput pipeline design.
With fine-grained pipeline execution, LightRW reduces the number of memory accesses to DRAM and achieves higher spatial parallelism compared to existing GDRW solutions.

We also propose two memory-efficient techniques to handle notoriously random accesses to the neighbors of vertices.
First, according to the power law distribution of graphs, vertices with high degrees dominate the connections and are frequently accessed.
Thus, we propose a degree-aware cache that keeps information of high-degree vertices (e.g., the addresses of their neighbors) into on-chip memory at runtime.
Second, we propose a dynamic burst access engine that adjusts the burst access size to minimize redundant data accesses and thus improve the use of available memory bandwidth, as the neighbors of different vertices have different lengths.

\subsection{Pipelining GDRW with WRS for FPGAs}\label{sec:adoptwrs}

To explore the solution of fully pipelined GDRW for FPGAs, we revisit existing weighted sampling methods.
We find that \textit{weighted reservoir sampling} (WRS) meets our requirements~\cite{efraimidis2006weighted} as it can choose $n_\textit{res}$ random samples from a population of unknown size in a single pass over the items.
In particular, let $w_i$ be the weight of the $i$-th item, and the probability that the $i$-th item is selected, denoted as $p_i$, is equal to its weight divided by the accumulated weights of all items passed, which is $p_i=\frac{w_i}{\sum_{m=1}^{i}w_m}$.
If $p_i$ is larger than a uniformly distributed random number $r_i$, the $i$-th item is then stored in a reservoir as a candidate for output, otherwise, it is discarded.
As GDRWs only need to sample one vertex to move forward, $n_\textit{res}$ is set to one.

The streaming processing nature of WRS eliminates the synchronization barrier between the initialization and generation stages required in the existing commonly used sampling methods, such as inverse transformation sampling and bucket-based sampling approaches~\cite{sun2021thunderrw,pandey2020c,hubschle2019parallel}.
However, it introduces a huge computational cost, as WRS requires a random number for each item of the input, and the generation of random numbers on CPUs is time-consuming.
This computational cost prevents WRS from being the best choice for the CPU-based GDRW system.
For example, when we adopt WRS in optimal CPU-based implementations, we observe that the performance is 8.2 times worse.
In contrast, generating massive random numbers on FPGAs is no longer a problem.
With a novel state sharing technique, ThundRiNG~\cite{thundering} is able to generate high-quality and high-throughput random numbers with efficient resource utilization on FPGAs.

\begin{algorithm}[t]
\SetInd{0.1em}{2em}
\KwData{$G$: a given graph,
$\mathbb{Q}$: a set of input queries.
}
\KwResult{$V_{t}$: a path of traversed vertices.}
\tikzmk{Zero}
$\textit{res} = \emptyset$\;
\ForEach{$Q\in \mathbb{Q}$}{
    $v_{\textit{curr}}= Q.v_{\textit{start}}$,
    $Q.\textit{res} = \emptyset$\;
\Loop{}{
    \tikzmk{Start}$\{\textit{address},\textit{degree}\} = \text{\textbf{get\_neighbors\_info}}(v_\textit{curr},G)$\;  \label{line:2_gni}\tikzmk{End}\boxline{myblue}
    \For{$i\gets1$ \KwTo $\textit{degree}$ \label{line:2_pqe_vertex_for}}
    {
        \tcc{weight\_calculation}
        \tikzmk{Start}
        $n_v = \text{\textbf{get\_neighbor}}(\textit{address},i , G)$\; \label{line:2_gn}
        \tikzmk{End}\boxfor{myblue}
        $w = \text{\textbf{app\_weight\_update}}(n_v, G)$\;\label{line:2_awu}

        \tcc{weighted\_sampling}
        \tikzmk{Start}{$v_\textit{curr} = \text{\textbf{WRS}}(n_,w, R)$}\; \label{line:2_pwrs}
    }
    \tikzmk{End}\boxline{myred}
    $Q.\textit{res}.\textbf{push}(v_\textit{curr})$\;
    \If{$(Q.{\normalfont \textbf{is\_end}}())$}
    {
        $\textit{res}.\textbf{push}(Q.\textit{res})$\;
        \KwBreak \;
    }
}
}
\caption{Dynamic Random Walk with WRS.}\label{alg:drw_move_one_step}
\end{algorithm}

In this paper, we adopt WRS for GDRWs to enable fine-grained pipeline execution on FPGAs, as shown in ~\Cref{alg:drw_move_one_step}.
Unlike CPU-based GDRWs shown in ~\Cref{alg:drw_nrs}, the weighted sampling method is changed to WRS.
This enables us to fuse all the neighbor weight update and sampling functions into one loop, and each function can produce and consume a single item at a time to allow fine-grained communication between functions.
We describe the detailed execution flow as follows:
First, {get\_neighbors\_info} loads $\{\textit{address},\textit{degree}\}$ of $v_\textit{curr}$ from graph $G$, where $\textit{address}$ is the location of $v_{curr}$'s neighbors in global memory, and \textit{degree} is the total number of $v_\textit{curr}$'s neighbors (\Cref{line:2_gni}).
Then, a for loop scans all $v_\textit{curr}$'s neighbors and randomly selects one item from them (~\Cref{line:2_pqe_vertex_for,line:2_gn,line:2_awu,line:2_pwrs}).
In the $i$-th iteration, it first loads the property of the $i$-th neighbor, $N_v$, and then calculates the sampling weight with {app\_update\_weight}, which is an application-specific function.
WRS takes the updated weight as input to temporally sample one item as $v_\textit{next}$.
After all the items are enumerated, the selected $v_\textit{next}$ is updated to $Q.v_\textit{curr}$ for the next step.

The intermediate variables of these functions, which are $n_v, w$, and $v_\textit{next}$, consume $O(1)$ space and are stored on-chip.
This is significantly smaller than the state-of-the-art execution flow required (i.e., $O(|N(v_\textit{curr})|)$~\cite{sun2021thunderrw}).
The proposed algorithm eliminates the synchronization barrier among different functions.
It also guarantees fine-grained pipeline execution on FPGAs, which reduces the number of memory accesses to DRAM by fine-grained communication on-chip and improves spatial parallelism by concurrently executing the computing stages.

\subsection{Hardware Architecture}\label{sec:overview}

LightRW efficiently realizes the proposed GDRW on FPGAs with a highly optimized microarchitecture.
While existing CPU-based GDRWs mainly focus on multi-query parallelization with task interleaving and multithreading, LightRW explicitly explores parallelism within the query by processing multiple neighbors per cycle.

\begin{figure}[t]
  \centering
  \includegraphics[width=\linewidth]{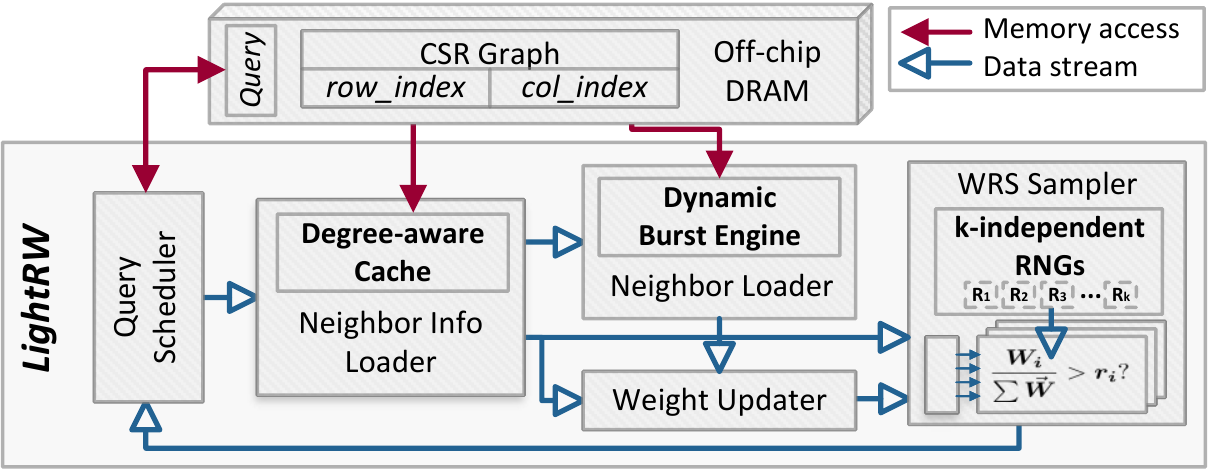}
     \caption{Overview of hardware architecture. }
     \label{fig:system_overview}

\end{figure}

\Cref{fig:system_overview} shows an overview of LightRW's hardware architecture, which consists of the \textit{query controller}, \textit{degree-aware cache}, \textit{neighbor info loader}, \textit{dynamic burst engine}, \textit{weight updater}, and \textit{WRS sampler}.
The target graphs are stored in the FPGA's DRAM with compressed sparse row (CSR) format, which consists of a \emph{row\_index} array and a \emph{col\_index} array.
The \emph{col\_index} array records adjacent edges (sorted by destination vertex) of each vertex, while the \emph{row\_index} records the offset of adjacent edges of each vertex in the \emph{col\_index} array.
The {Query Controller} loads multiple queries and prepares query metadata (e.g., the $v_{curr}$, query index, and application-specific parameters) for the next stage.
The {Neighbor Info Loader} reads the address of the neighbors (adjacent edges) of $v_{curr}$, namely the \emph{row\_index} of $v_{curr}$.
As $v_{curr}$ is randomly selected, the accesses to the addresses of neighbors are random.
Our proposed \emph{Degree-aware Cache} caches vertices with high degrees to improve memory efficiency (details in \ref{sub:degree_aware_cache}).
The {Neighbor Loader} sequentially loads the neighbors of $v_{curr}$ by dereferencing the address of neighbors.
However, the number of neighbors varies according to the vertices. The memory coalescing with a fixed burst length may result in redundant memory accesses.
Therefore, we also propose a \emph{Dynamic Burst Engine} that adopts hybrid burst lengths to maximize memory bandwidth utilization (details in \ref{sub:dynamic_burst_engine}).
The {Weight Updater} calculates the sampling weight of the fetched neighbors using an application-specific update function. The \emph{WRS Sampler} consumes multiple neighbors per cycle with throughput at the line rate of memory bandwidth and samples the vertex for the next step of the current query.
Details on how to parallelize WRS are given in Section~\ref{sec:parallelWRS}.
The {Query Controller} updates $v_{curr}$ with the sampled vertex.
All the above hardware modules are pipelined and run concurrently.

\section{Parallelizing WRS}
\label{sec:parallelWRS}

\begin{figure*}[t]
  \centering
  \includegraphics[width=\linewidth]{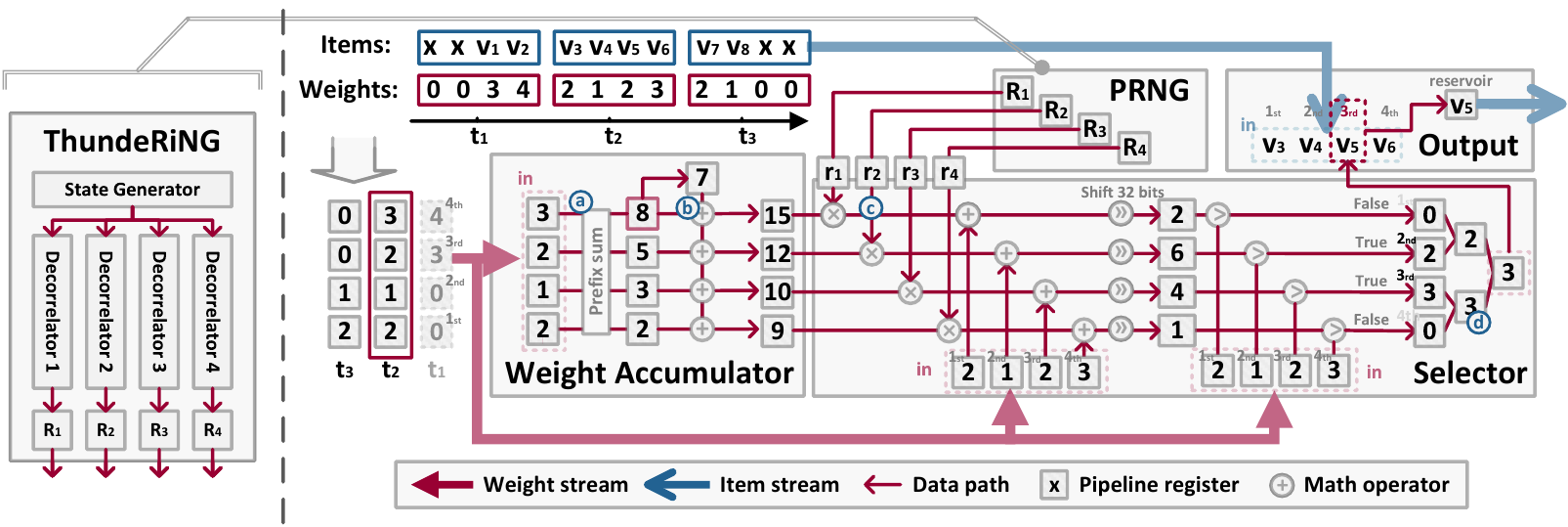}
     \caption{\rv{Hardware architecture of WRS Sampler.}}
     \label{fig:hwrs}
     \vspace{2mm}
\end{figure*}

To process multiple neighbors per cycle for GDRWs, WRS has to be parallelized. However, data dependency in calculating the probability of the item in the reservoir prevents parallelization.
As discussed in~\Cref{sec:adoptwrs}, the probability of the $i$-th item, $p_i$, is calculated as its weight divided by the accumulated weight of all passed items, i.e., $ p_i = \frac{w_i}{\sum_{m=1}^{i}w_m}$.
In other words, the probability calculation of the current item is dependent on the weights of the previously processed items, which prevents straightforward parallelization of the probability calculation. For instance, simple loop unrolling cannot be applied due to this dependency~\cite{huang1999generalized}.
Existing works on paralleling WRS~\cite{jayaram2019weighted,hubschle2019parallel} aim to reduce the number of required random numbers for each machine in a distributed environment, as random number generation is usually time-consuming for CPUs. 
On the contrary, FPGAs can easily generate multiple independent random numbers benefiting from spatial parallelism, high-performance bitwise operations, and flexible on-chip communication~\cite{thundering}. Therefore, in this paper, we propose a new parallelized WRS algorithm and its FPGA-based implementation.

\subsection{Parallel WRS Algorithm}

The key idea of our proposal is to consume multiple items per cycle while carefully handling their dependency to ensure the independence and correctness of sampling.
As each item in the stream requires the accumulated weight of all items passed, namely $w_\textit{sum}^i = \sum_{m=1}^{i}w_m$, we employ a parallelized prefix sum to calculate the accumulated weights for multiple items in parallel.
Specifically, assuming that we process $k$ items per cycle and $i$ items have passed, the accumulated weights for the current $k$ items can be represented by the set $\{\sum_{m=1}^{i+j} {w_m}\}_{j = 1}^{k}$.
By decomposing the sum of the weights of the passed $i$ items, we have the following representation of the set $\{\sum_{m=1}^{i+j} {w_m}\}_{j = 1}^{k}$,
\begin{align}
\{\sum_{m=1}^{i+j} {w_m}\}_{j = 1}^{k}
&= \{\sum_{m=1}^i {w_m} + \sum_{m = i+1}^{i+j}{w_m}\}_{j = 1}^{k}.
\end{align}
As $w_\textit{sum}^i =\sum_{m=1}^i {w_m}$,  which is a constant for current $k$ items, we have
\begin{align}
\{\sum_{m=1}^{i+j} {w_m}\}_{j = 1}^{k}
&= \{w_\textit{sum}^{i} + \sum_{m = i+1}^{i+j}{w_m}\}_{j = 1}^{k}\\
&=w_\textit{sum}^{i} +\{\sum_{m = i+1}^{i+j}{w_m}\}_{j = 1}^{k},\label{eq:perfix_sum} 
\end{align}
where $\{\sum_{m = i+1}^{i+j}{w_m}\}_{j = 1}^{k}$ is the prefix sum of $\{w_{i+1}, w_{i+2},...w_{i+k}\}$ that can be calculated in parallel.

\begin{algorithm}[t]
\SetInd{0.1em}{2em}
\KwData{ \begin{alignat*}{4}
&{N_v}=\{n_{v_1},n_{v_2},...\} &&: \text{item stream to be sampled;}\\
&{W}=\{w_1,w_2,...\} &&: \text{corresponding weight stream;} \\
&k &&: \text{degree of parallelism;}\\
&{R}=\{r_1,r_2,...,r_k\} &&: \text{k-wise independent random variables.}
\end{alignat*}}
\KwResult{ \begin{alignat*}{4}
n_{s}~~~~~~~~~~~~~~~~~~~~~~~~: \text{item sampled with probability of}\frac{w_{s}}{\sum{W}}.
\end{alignat*}}

$w_\textit{sum}^i = 0$, $\textit{reservoir} = \emptyset$, $n_{s} = \emptyset$ \;
\Do{ $N_v$ is end}{
\tcc{get $k$ items and corresponding weights from streams}

 \If{$(\hat{N_{k}} ={\normalfont \textbf{receive}}(N_v,k)) \land (\hat{W}_k = {\normalfont \textbf{receive}}(W,k))$ \label{line:3_receive}}{
\tcc{calculate the prefix sum array of $\hat{W}_k$}
$W_\textit{ps} = \textbf{prefix\_sum}(\hat{W}_k)$\; \label{line:3_perfix_sum}
$\textit{candidate} = \emptyset $\; 
\Pfor{$j\gets1$ \KwTo $k$ \label{line:3_pfor}}{
    \tcc{calculate the probability of selecting $\hat{W}_k[j]$}
     $p =  \frac{\hat{W}_k[j]}{w_{sum}^i + W_\textit{ps}[j]}  $\; \label{line:3_probability}
     $r = R[j].\textbf{sample}(0,1)  $\; \label{line:3_sample}
     
    \If{p > r \label{line:3_if}}{
        $\textit{candidate}[j] = j$\; \label{line:3_mask_set}
    }
}
\tcc{find the max index in $\textit{candidate}$}
$\textit{sel} = \textbf{max}(\textit{candidate})$\; \label{line:3_sel}
\tcc{update the reservoir}
    \If{$sel \neq 0$}{
        $\textit{reservoir} = \hat{N_k}[\textit{sel}] $\;
    }
\tcc{update $w_\textit{sum}^i$ value for next $k$ items}
$w_\textit{sum}^i = w_\textit{sum}^i +  \textbf{sum}(\hat{W}_k)$ \label{line:3_sum}\;
 $n_{s} = \textit{reservoir}$ \;
\Output{$n_{s}$}\;
}
}
\Return{$n_{s}$}
\caption{Parallel Weighted Reservoir Sampling.}\label{alg:pwrs}
\end{algorithm}

\Cref{alg:pwrs} shows the proposed parallelized WRS that can process $k$ items per cycle.
The inputs consist of the item stream ${N_v}=\{n_{v1},n_{v2},...\}$, weight stream ${W}=\{w_1,w_2,...\}$, the degree of parallelism $k$, and $k$-wise independent random variables uniformly distributed in the interval $[0,1]$, ${R}=\{R_1,R_2,...,R_k\}$.
The algorithm reads $k$ items and the corresponding weights per cycle (\Cref{line:3_receive}).
\Cref{line:3_perfix_sum} calculates the prefix sum of the $k$ items, which corresponds to the second term $\{\sum_{m = i+1}^{i+j}{w_m}\}_{j = 1}^{k}$ in~\Cref{eq:perfix_sum}, and stores it in the array $W_{ps}$. 
Then, $k$ items are sampled concurrently. For each item in $\hat{N_{k}}$, \Cref{line:3_probability} calculates the probability of the $j$-th item being selected. Next, a random number $r$ is sampled from the $j$-th random variable in $R$ (Line 8).
If $p$ is smaller than $r$, its index is stored in $\textit{candidate}$ (Lines 9 to 10), indicating that the $j$-th item is temporarily selected.
After all $k$ items are processed, the maximum value in $\textit{candidate}$, which is the index of the latest sampled item, is outputted to update the $\textit{reservoir}$ (Lines 11 to 13).
Meanwhile, $w_\textit{sum}^i$ is accumulated with $\emph{sum}(\hat{W}_k)$ (\Cref{line:3_sum}) for the next batch of sampling.
For a stream with $n$ items, the time complexity of the proposed algorithm is $O(n/k + \log{k})$, where the functions \emph{prefix\_sum()}, \emph{max()}, and \emph{sum()} introduce the $O(\log{k})$ time complexity.
Since only $\textit{reservoir}$ needs to be stored for each $k$ input, the space complexity of the proposed algorithm is $O(1)$.

\subsection{WRS Sampler} %
\label{subsec:wrs_sampler}
\rv{Deploying our proposed parallel WRS algorithm on FPGAs highly relies on high-quality and high-throughput random number generation (RNG). Therefore, we first select a suitable RNG module. After that, we introduce the hardware architecture of the WRS Sampler module for our proposed parallel WRS algorithm.
}

\noindent\textbf{Selection of RNGs.}
\rv{
RNGs can be classified into two types: true random number generations (TRNGs)\cite{stipvcevic2014true} and pseudo-random number generations (PRNGs)\cite{matsumoto1998mersenne}. TRNGs produce unpredictable and unreproducible outputs, while PRNGs are commonly used in randomized algorithms such as Monte Carlo simulations~\cite{rubinstein2016simulation} and graph random walks~\cite{rwprinciples} to ensure the reproducibility of results. In recent years, FPGA-based PRNGs have gained attention from both industry and academia~\cite{li2011software,bakiri2018survey}.
}

\rv{To enable high-performance WRS, we have summarized the following criteria for selecting PRNGs:
First, the PRNGs must be capable of generating multiple independent sequences of random numbers to ensure the independent selection of the candidates in~\Cref{line:3_pfor,line:3_probability,line:3_sample,line:3_if,line:3_mask_set} of~\Cref{alg:pwrs}.
Second, the generation of random numbers should be fast enough to avoid blocking the overall processing pipeline.
Third, the hardware resources required to construct the PRNGs should not constitute a severe conflict with respect to the resource demand for GDRW processing.
}

\rv{
To meet all of the aforementioned criteria, we have selected ThunderRiNG~\cite{thundering} and integrated it into our WRS pipeline, as shown in Figure~\ref{fig:hwrs}. ThunderRiNG allows for the generation of multiple independent sequences of random numbers while sharing the costly state generation among different instances. Additionally, each instance adopts a decorrelator to generate a high-quality random number sequence. ThunderRiNG also provides an easy-to-use interface for integration. Our evaluation has demonstrated that ThunderRiNG is capable of simultaneously generating up to 64 sequences of random numbers, consuming only $1.2\%$ of hardware resources, and passing the most stringent empirical randomness tests~\cite{l2007testu01}.
}

\noindent\textbf{WRS Sampler Hardware Architecture.}
~\Cref{fig:hwrs} illustrates the hardware architecture of the WRS Sampler that realizes the proposed parallelized WRS. 
It includes four hardware modules: \emph{Weight Accumulator}, \emph{Selector}, \emph{PRNG}, and \emph{Output}.
To facilitate presentation, we set $k$ to four.
The \emph{Weight Accumulator} reads $k$ items from the weight stream in one cycle (e.g.,~$\{2,1,2,3\}$ at time $t_2$) and calculates the accumulated weights of $k$ items in two steps.
First, it calculates $W_{ps}$ using pipelined prefix sum logic (Step \circled{a}). Then, in Step \circled{b}, it calculates $\{w_\textit{sum}^i + W_\textit{ps}[j]\}_{j=1}^k$ using four adders in parallel, and the last prefix sum (e.g., eight in the figure) is added to $w\textit{sum}$ to calculate the next $k$ items. The \emph{Selector} performs the sampling procedure of $k$ items per cycle. To avoid costly division operations in calculating $p$, we rewrite the comparison between $p$ and the random number in the interval of [0, 1], $r$, as follows: First, let $r*$ be a 32-bit integer random number generated from ThundeRiNG. The condition for sampling the $j$-th item is as follows,
\begin{align}
p > r \Rightarrow \frac{{\hat{W}_k}[j]}{w_{sum}^i + W_{ps}[j]} > \frac{r^*}{2^{32} -1}.
\end{align}
Then, by moving the denominators from both sides to the numerators on the other side, we obtain the following inequality:
\begin{align}
\Rightarrow  {\hat{W}_k}[j](2^{32} - 1) > r^* (w_{sum}^i + W_{ps}[j]).
\end{align}
We further reorganize the left term,
\begin{align}
\Rightarrow  2^{32}\cdot{\hat{W}_k}[j] > r^* (w_{sum}^i + W_{ps}[j]) + {\hat{W}_k}[j].
\label{eq:reform}
\end{align}
The comparison in ~\Cref{eq:reform} can be further simplified as follows.
Multiplication in the left term can be obtained by light-weight bit shifting on FPGAs.
As $\{w_\textit{sum}^i + W_\textit{ps}[j]\}_{j=1}^k$ is the output of the Weight Accumulator, the right term only needs one multiplication and one addition (Step \circled{c}), which can be completed simultaneously by the DSP slice in FPGAs~\cite{rodriguez2007features}.
If the comparison condition is true, which means the item is a candidate for output, we record the indices of these items (e.g., two and three in this example).
Finally, we determine the latest candidate by comparing their indices.
This is implemented using a tree-based comparator (Step~\circled{d}), where each level compares two adjacent items. Based on the index of the selected item, \emph{Output} extracts the corresponding item from the input item stream and updates $\textit{reservoir}$. In the example, the third item, $v_5$, is selected in the $\textit{reservoir}$. When the stream ends, $\textit{reservoir}$ is output as the sampled result.

\section{Memory Optimizations}
\label{sec:memory_optimizations}

Memory accesses to DRAM are highly optimized in LighRW. First, LightRW employs a degree-aware cache that stores only the high-degree vertices in on-chip memory. Second, since neighbors of different vertices have varying lengths, we propose a dynamic burst access engine that adjusts the burst access size to minimize redundant data access, thereby improving memory bandwidth utilization.

\subsection{Degree-aware Cache} %
\label{sub:degree_aware_cache}
Accesses to the addresses of neighbors of the current vertex, namely the accesses to the \emph{row\_index} array in the Neighbor Info Loader component, are inherently random, as the current vertices from different steps are selected randomly.
Due to the large reuse distance~\cite{basak2019analysis}, existing cache policies (i.e., LRU and FIFO, etc.) that are designed to retain the most recent or frequently-used data items in on-chip memory are ineffective in handling this access.
In this paper, we provide an analysis of how the degree of a vertex can indicate the reuse ratio of a vertex in GDRWs. Based on this analysis, we propose a degree-aware cache to cache the high-degree vertices at runtime with zero initialization overhead.

Buffering high-degree vertices into fast memory for graph processing has been adopted in many recent research works~\cite{zhao2021lccg,arai2016rabbit,balaji2019combining}. These approaches are based on the observation that graph processing applications frequently access the properties of vertices with a larger number of neighbors. For instance, \citeauthor{zhao2021lccg}\cite{zhao2021lccg} build a hash lookup table for high-degree vertices during graph partitioning, while~\citeauthor{balaji2019combining}~\cite{balaji2019combining} sort the vertices by their degrees in the preprocessing phase and then reindex all vertices during the graph data preprocessing. However, all of these existing works introduce additional initialization overhead and are only applicable for graph processing. Instead, we explore the degree-aware caching method for GDRWs.
\begin{figure}[t]
     \centering
     \includegraphics[width=\columnwidth]{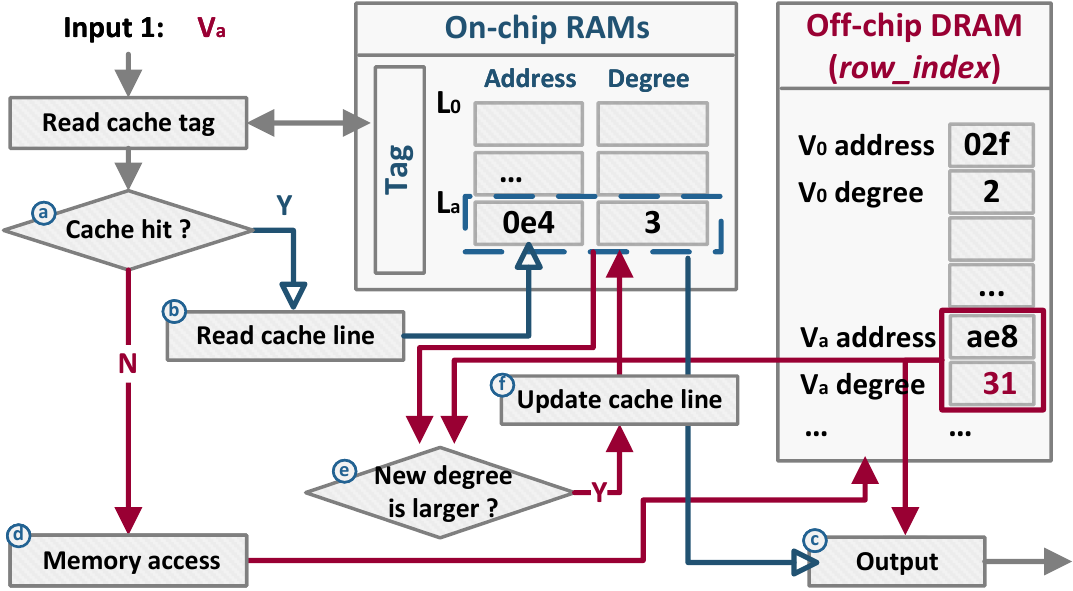}
     \vspace{3pt}
     \caption{Hardware architecture of degree-aware cache.}
     \label{fig:dac}
     \vspace{-2pt}
\end{figure}

\noindent\textbf{Probability Analysis of Vertex Accesses:} We begin by presenting the analysis that the degree of a vertex has a positive correlation with the probability of the vertex being traversed in GDRWs, by extending the existing conclusion on static random walks. Let $Pr[\text{$v$}]$ be the probability that $v$ is traversed by multiple independent random walk queries on a static weighted graph and follows a stationary distribution~\cite{rwprinciples}. Then, $Pr[\text{$v$}]$ can be represented as follows:
\begin{align}
Pr[\text{$v$}]  = \dfrac{\sum_{u \in N(v)} w_{v,u}}{\sum_{i\in V} \sum_{j \in N(i)} w_{i, j}}.
\end{align}

\noindent Next, we consider dynamic random walks, where the sampling weight from vertex $u$ to vertex $v$ at time $t$ is $w_{u, v}^t$. The stationary distribution of $v$ based on the sampling weights at time $t$ is then given by
\begin{align}
Pr[\text{$v$}]  = \dfrac{\sum_{u \in N(v)} w_{v,u}^t}{\sum_{i\in V} \sum_{j \in N(i)} w_{i, j}^t}.
\end{align}
Let $\mathcal{W}$ be defined as $\sum_{i\in V} \sum_{j \in N(i)} w_{i, j}^t$, and let $\alpha$ be a scaling factor that scales the minimal non-zero weight to one. Since graphs have a finite number of edges, $\alpha$ exists and is constant for a given graph. Similarly, $\mathcal{W}$ is also a constant value. Therefore, we have:
\begin{align}
 Pr[\text{$v$}]  &= \frac{\sum_{u \in N(v)} \alpha w_{v,u}^t}{\alpha \mathcal{W}}.
\end{align}
Meanwhile, there exists a scaling factor $\alpha$ such that the numerator $\sum_{u \in N(v)} \alpha w_{v,u}^t$ is larger than $N(v)$ and holds for any $t$. Since $1/{\alpha \mathcal{W}}$ is a constant value, we can conclude that $Pr[\text{$v$}] = \Omega(N(v))$. Hence, the degree of vertices can be an admissible heuristic that estimates the probability of the corresponding vertex being traversed.

\noindent\textbf{Degree-aware Cache Policy:} The above analysis supports us in designing a cache replacement policy that replaces low-degree vertices in fast memory with high-degree vertices for a high cache hit ratio.
\Cref{fig:dac} shows an example of our cache replacement policy and the hardware architecture of the proposed degree-aware cache.
The input to our degree-aware cache is a vertex index, and the corresponding output is a tuple that contains the starting address of the neighbors (adjacent edges) and the degree of the vertex.
When an input vertex $v_a$ is received in Step~\circled{a}, our degree-aware cache looks up whether $v_a$ exists in the cache by comparing the tag of the cache line and the tag of the input.
If $v_a$ is already in the cache (cache hit), the corresponding cache line is read in Step~\circled{b}, and the output is directly sent to the Neighbor Loader module within one cycle (Step~\circled{c}).
If $v_a$ is not in the cache (cache miss), one memory access is issued to the off-chip DRAM in Step~\circled{d}.
When the data for $v_a$ is received from DRAM, the cache immediately outputs the returned data and compares the degree of the vertex in the corresponding cache line with the degree of $v_a$ in Step~\circled{e}.
If the degree of $v_a$ is higher, the cache updates the cache line with the data for $v_a$; otherwise, it retains the original data.

\subsection{Dynamic Burst Engine} %
\label{sub:dynamic_burst_engine}

\begin{figure}[t]
     \centering
     \includegraphics[width=\columnwidth]{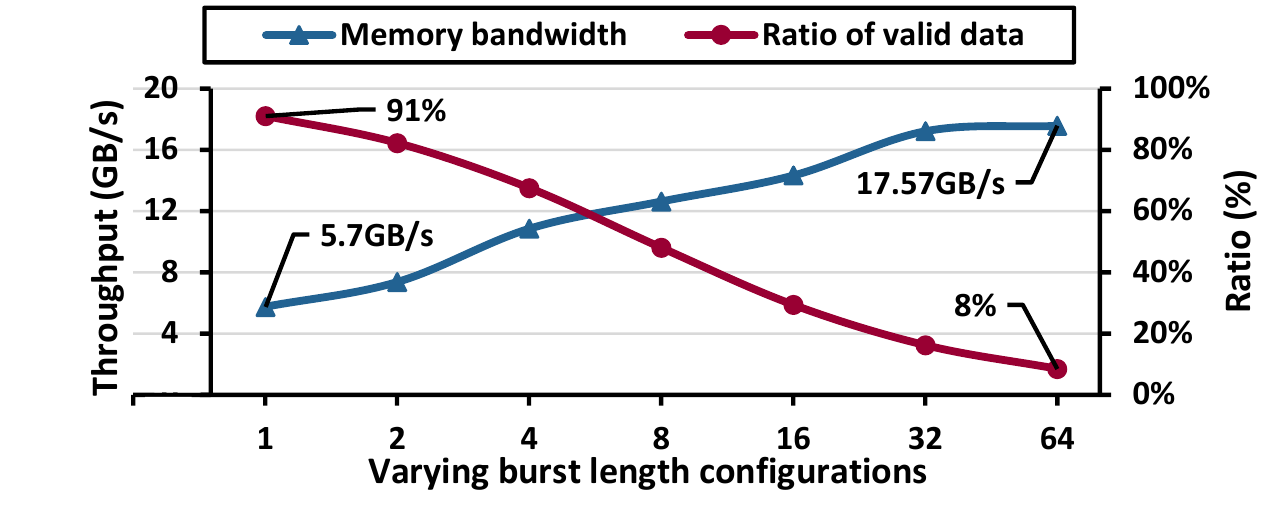}
     \caption{Memory bandwidth and the ratio of valid data of MetaPath on livejournal with varying burst lengths.}
     \label{fig:burst_eva}
     \vspace{-3pt}
\end{figure}

GDRWs need to load all neighbors of the current vertex to move one step forward.
Since neighbors of a vertex are stored in consecutive addresses, burst access that reads multiple data sequentially with a single memory request has been an effective method to improve memory bandwidth utilization~\cite{chen2022thundergpr,choi2021hbm, 8892073}.
The data size of each burst access, namely the burst size, is equal to the data width of the memory interface, $N_\textit{bus}$, multiplied by the burst length, $S$. For example, if the data width of the memory interface is 512-bit, burst access with a burst length of four will request 2048-bit data per request.
\Cref{fig:burst_eva} shows memory throughput benchmark results (blue line) with different burst lengths on our FPGA platform (\Cref{sub:experimental_setup}). 
The results indicate that the utilized memory bandwidth increases significantly with increasing burst length until reaching the peak bandwidth ($17.57$ GB/s).

\noindent\textbf{Issues of fixed burst length:} Burst access with a long burst length is not effective for power-law graphs as vertices have varying numbers of neighbors.
If the size of the neighbors is narrower than the burst size, the remaining data returned from memory will be unused, resulting in poor effective memory bandwidth utilization.
We define \emph{the ratio of valid data} as the percentage of data used in the computation compared to the total number of data loaded by the memory engine.
The red line in \Cref{fig:burst_eva} shows the ratio of valid data of MetaPath on livejournal with an increasing burst length configuration. It turns out that the ratio of valid data is the highest with a burst length of one and decreases with larger burst lengths.
The existing study, DynaBurst proposed by~\citeauthor{8892073}\cite{8892073}, adopts variable-length bursts to the memory interface to improve the effective memory bandwidth for irregular applications. The main idea is to cache and reorder numerous narrow memory requests in a non-blocking cache and explore data reuse of burst accesses.
However, it requires a large amount of on-chip memory and logic resources for the cache and reordering function, respectively. Moreover, the latency is non-deterministic and varies significantly, which is not suitable for GDRWs.

\noindent\textbf{Our Design:} \rv{\setword{To}{rv:r1o1}  utilize the high memory bandwidth of the long burst and maintain a high ratio of valid data of the short burst simultaneously, we propose a dynamic burst engine that schedules memory access requests with varying sizes to access pipelines with different burst length configurations at runtime.}
Let $S_1$ and $S_2$ be the number of bytes that can be loaded by one burst access of the long burst pipeline and short burst pipeline, respectively. With a total of $c$ bytes of neighbors to be loaded, the number of long burst accesses is set to $\floor{c/S_1}$, while the number of short bursts is set to $\ceil{(c - \floor{c/S_1}S_1)/S_2}$.
In this way, the majority of requested data is handled by the long burst pipeline with high memory bandwidth utilization, and the remaining data that cannot form a long burst is loaded by the short burst pipeline, achieving a high ratio of valid data. As the total bytes loaded is equal to $\floor{c/S_1}S_1 + \ceil{(c - \floor{c/S_1}S_1)/S_2}S_2$, which is equal to $\ceil{c/S_2}S_2$, the loaded unused data is no larger than $S_2$.

\begin{figure}[t]
\begin{minipage}[t]{.45\textwidth}
 \includegraphics[width=\linewidth]{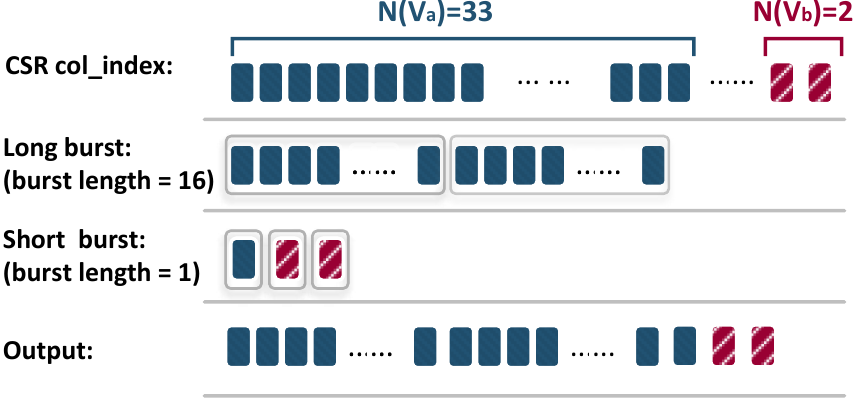}
    \caption{Example of dynamic burst strategy.}
    \label{fig:dyb_pg}
\end{minipage}\hfill
\begin{minipage}[t]{.5\textwidth}
  \includegraphics[width=\columnwidth]{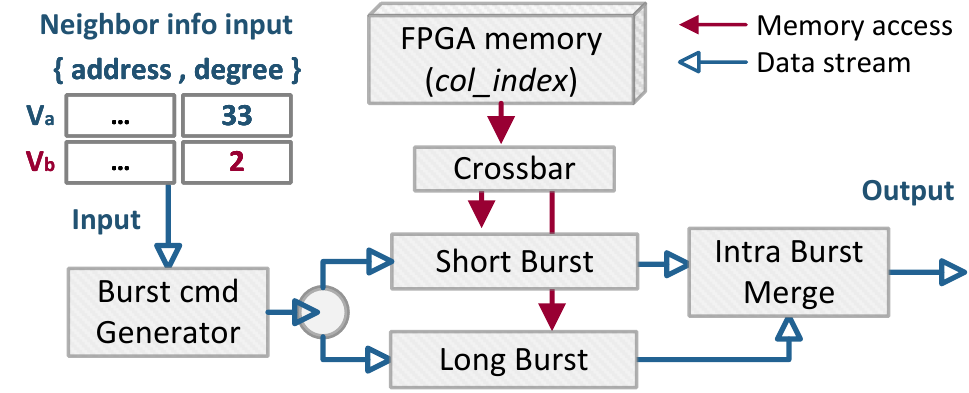}
     \caption{Hardware architecture of dynamic burst engine.}
     \label{fig:dyb_ha}
\end{minipage}
\end{figure}

\rv{Due to the highly skewed distribution of node degrees in real graphs, it is very hard to build the model of burst length configurations to graphs with different structures. Hence, we determine the configurations of the burst length from practical evaluation (more details in~\Cref{sub:dybee}).}
\Cref{fig:dyb_pg} presents an example of our dynamic burst strategy given $S_1 = 16$ and $S_2=1$. Assuming the request to neighbors with a size of 33, the request is scheduled into two ($\floor{33 / 16}$) long burst accesses and one ($(33 - 2\times16)/1$) short burst access.
The request to neighbors with a size of two consists of zero ($\floor{2 / 16}$) long burst and two ($(2 -0)/1$) short burst accesses.

\Cref{fig:dyb_ha} presents the hardware architecture of our dynamic burst engine, which contains four hardware modules: Burst cmd Generator, Long Burst, Short Burst, and Intra Burst Merge. 
The Burst cmd Generator module takes a data pair, ${\textit{address}, \textit{degree}}$, as input, where \textit{address} is the starting address of neighbors in \emph{col\_index}, and \textit{degree} is the number of neighbors of the vertex.
The Burst cmd Generator follows the above scheduling method to generate burst access commands and dispatches commands to the Long Burst and Short Burst modules.
Long Burst and Short Burst are connected to a memory crossbar and access the \emph{col\_index} array in FPGA's DRAM independently.
Their burst length is configured to $S_1$ and $S_2$, respectively. The Intra Burst Merge module returns data from two burst modules and outputs them to the Weight Updater.

\section{Evaluation}

\label{sec:evaluation}
In this section, we present the evaluation of LightRW.
In~\Cref{sub:experimental_setup}, we introduce the hardware setup and the graph dataset used in our evaluation.
In~\Cref{sub:eva_pwrs}, we evaluate the throughput and efficiency of the proposed weighted reservoir sampling module, PWRS sampler. We further demonstrate the impact of parameters in the dynamic burst engine and degree-aware cache in~\Cref{sub:memory_opt}.
Finally, we compare LightRW with the state-of-the-art CPU-based implementation in~\Cref{sub:comparison_with_state_of_the_art_cpu_imp}.

\begin{figure}[h]
     \centering
     \includegraphics[width=\columnwidth]{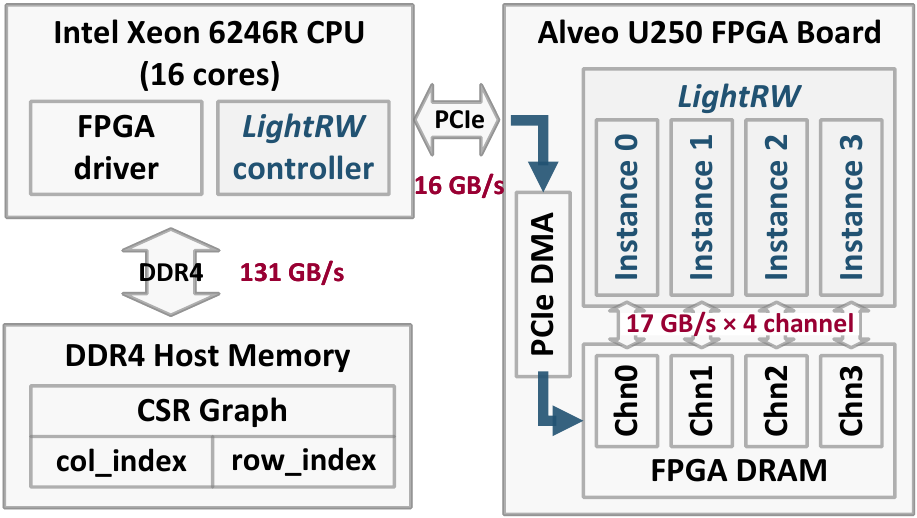}
     \caption{Deployment of LightRW.}
     \label{fig:deployment}
\end{figure}

\subsection{Experimental Setup}
\label{sub:experimental_setup}
\subsubsection{Hardware Platform}
LightRW is built on a machine where an Xilinx Alveo U250 accelerator card is attached to the motherboard via the PCIe bus, as shown in~\Cref{fig:deployment}. The FPGA board has 2,000 BRAMs, 11,508 DSP slices, 1,341,000 LUTs, and four DRAM channels with a total capacity of 64 GB. We used the Vitis HLS Toolchain 2021.2 as the development environment.

\subsubsection{Graph Datasets}
\Cref{tab:graphdataset} shows the graph datasets used in our evaluation. These real-world graphs come from different categories, including the web, citations, and social networks. The RMAT synthetic graphs are produced by the RMAT generator~\cite{chakrabarti2004r}.

\begin{table}[h]
\centering
\caption{The graph datasets.}
\label{tab:graphdataset}
\resizebox{1\linewidth}{!}{%
\begin{threeparttable}
\begin{tabularx}{1.25\linewidth}{lrrrrr}
\toprule
\multicolumn{1}{l|}{Graphs}  & \text{$\left| V \right|$} &  \text{$\left| E \right|$}&\text{$\left| D \right|$} & Type & Categories \\
\midrule

\multicolumn{1}{l|}{ youtube (\emph{YT})  ~\cite{snapnets}} & 1.14M & 2.99M & 5 & Undirected & Web \\
\multicolumn{1}{l|}{ us-patents (\emph{UP})  ~\cite{snapnets}} & 3.78M & 16.52M & 9 & Directed & Citation \\
\multicolumn{1}{l|}{ liveJournal (\emph{LJ})  ~\cite{snapnets}} & 4.8M & 68.9M & 14 & Undirected & Social \\
\multicolumn{1}{l|}{ orkut (\emph{OR})  ~\cite{graphdataset}} & 3.1M & {117.2M} & 38  & Undirected & Social \\
\multicolumn{1}{l|}{ uk2002 (\emph{UK})  ~\cite{BoVWFI}} & 18.52M & {298.11M} & 32  & Directed & Social \\
\multicolumn{1}{l|}{ rmat-$12\unsim22$ (\emph{RMAT}) ~\cite{kronecker2010}} & $2^{12}\unsim2^{22}$ & $2^{15}\unsim2^{25}$ & 8 & Directed & Synthetic  \\

\bottomrule
\end{tabularx}%

\end{threeparttable}
}

\end{table}

\subsubsection{Workloads}
We implement and evaluate two representative GDRW algorithms: MetaPath~\cite{sun2013mining} and Node2Vec~\cite{grover2016node2vec}. These algorithms are widely adopted as benchmarks in existing graph random walk systems~\cite{grover2016node2vec,dong2017metapath2vec,nikolentzos2020random,10.1145/3079628.3079685}.

\subsubsection{Parameter Settings}
\label{ssub:parameter_settings}
We follow the experimental settings of previous works~\cite{sun2021thunderrw,yang2019knightking}. The number of queries is set to the number of vertices with non-zero degrees, and each query has a unique starting vertex. All queries are shuffled following the approach in ThunderRW~\cite{sun2021thunderrw}. The sampling method of ThunderRW is configured to use inverse transformation sampling following the authors' recommendation. The graph datasets are initialized with random edge weights and vertex labels. The query length is set to five for MetaPath and 80 for Node2Vec, and we set $p = 2$ and $q = 0.5$ for Node2Vec. All experiments are repeated five times, and the median results are presented.

\subsubsection{Implementation Details}\label{subsec:implementation}
As shown in~\Cref{fig:deployment}, we implemented LightRW on a CPU-FPGA platform, where a Xilinx Alveo U250 FPGA accelerator card is attached to the motherboard through the PCIe bus. To run GDRW queries on a given graph, the LightRW controller on the host CPU side issues DMA requests to transfer random walk queries and graph data in the compressed sparse row (CSR) format to the DRAM of the FPGA platform and then invokes the accelerator. The LightRW controller then enters a waiting state until the accelerator completes the computation.

Modern FPGAs generally use multiple DRAM channels to provide higher memory bandwidth~\cite{korolija2020abstractions,chen2021thundergp}.
To utilize multiple DRAM channels, we instantiate multiple independent LightRW instances and connect each of them to one DRAM channel, as shown in~\Cref{fig:deployment}.
Each LightRW instance has a private and independent copy of the graph data and is configured to fully utilize the bandwidth of one channel.
Meanwhile,  we evenly distribute random walk queries to all instances.

\subsection{Evaluation on WRS Sampler}
\label{sub:eva_pwrs}

\begin{figure}[t]
\begin{minipage}{\linewidth}
  \subcaptionbox{Varying degree of parallelism.\label{fig:pwrs_k}}
    {\includegraphics[width=0.49\linewidth]{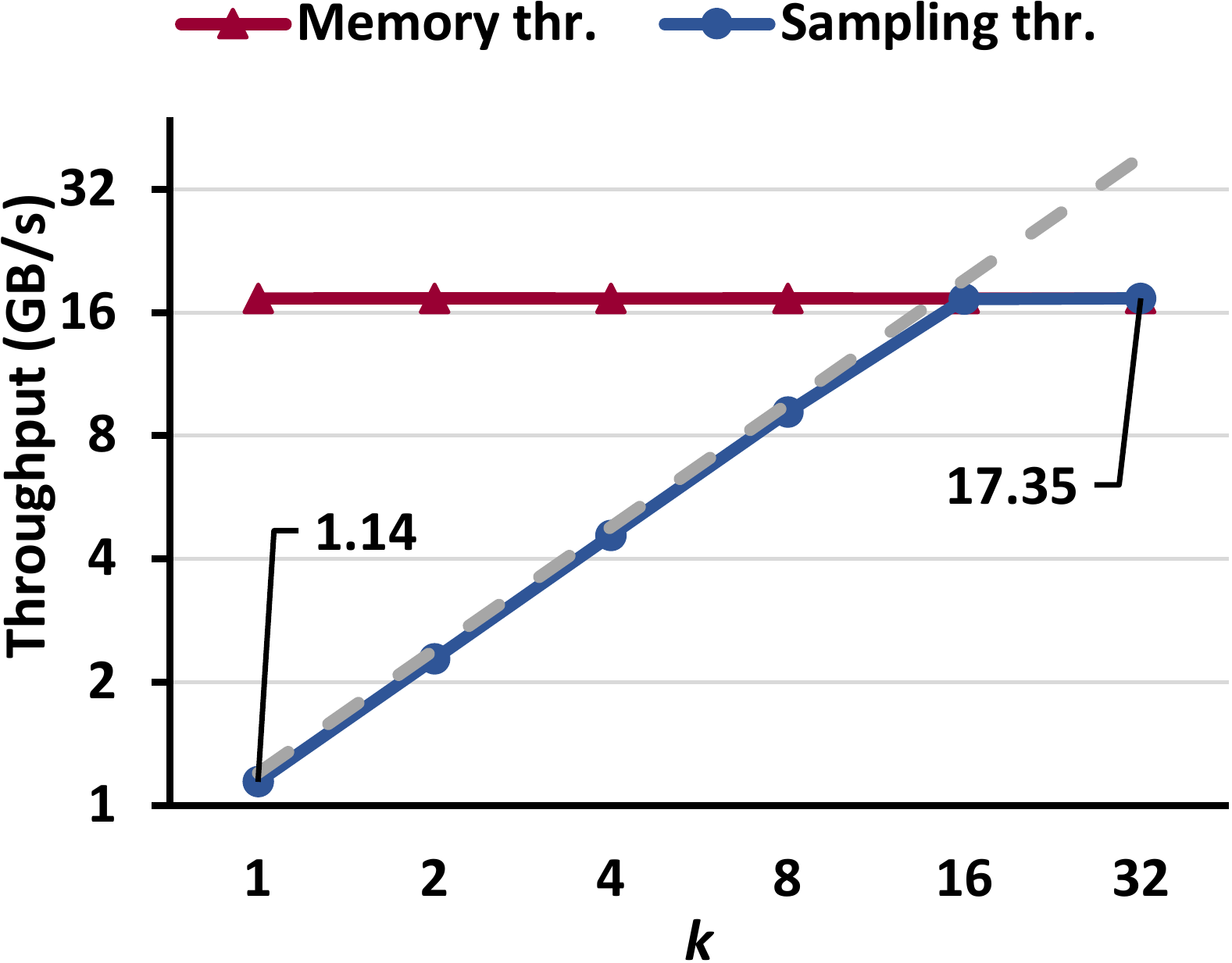}}
  \subcaptionbox{Varying length of stream.\label{fig:pwrs_size}}
    {\includegraphics[width=0.49\linewidth]{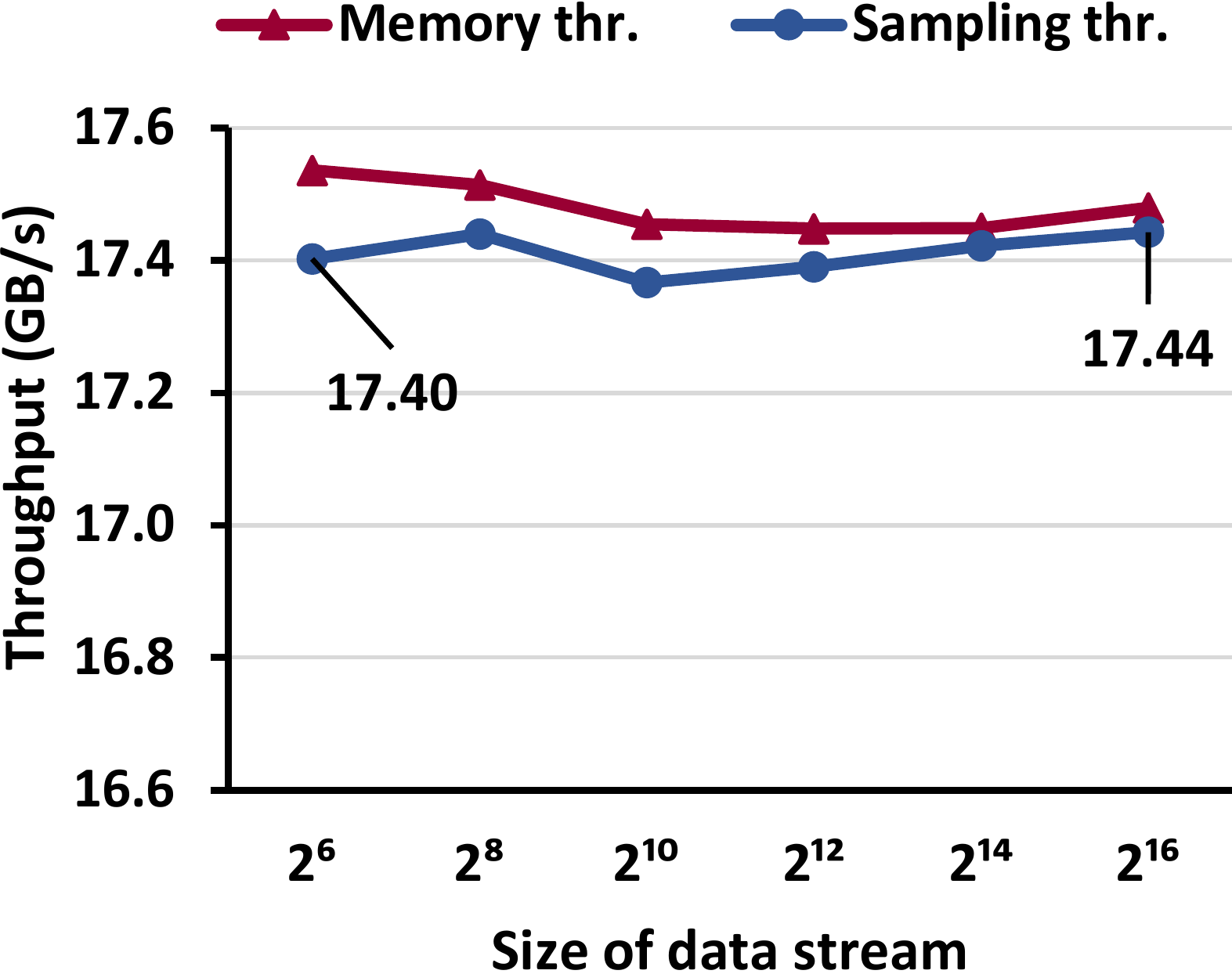}}

  \caption{Throughput evaluation of WRS sampler.}
\end{minipage}

\begin{minipage}{\linewidth}
\centering
  \includegraphics[width=0.7\linewidth]{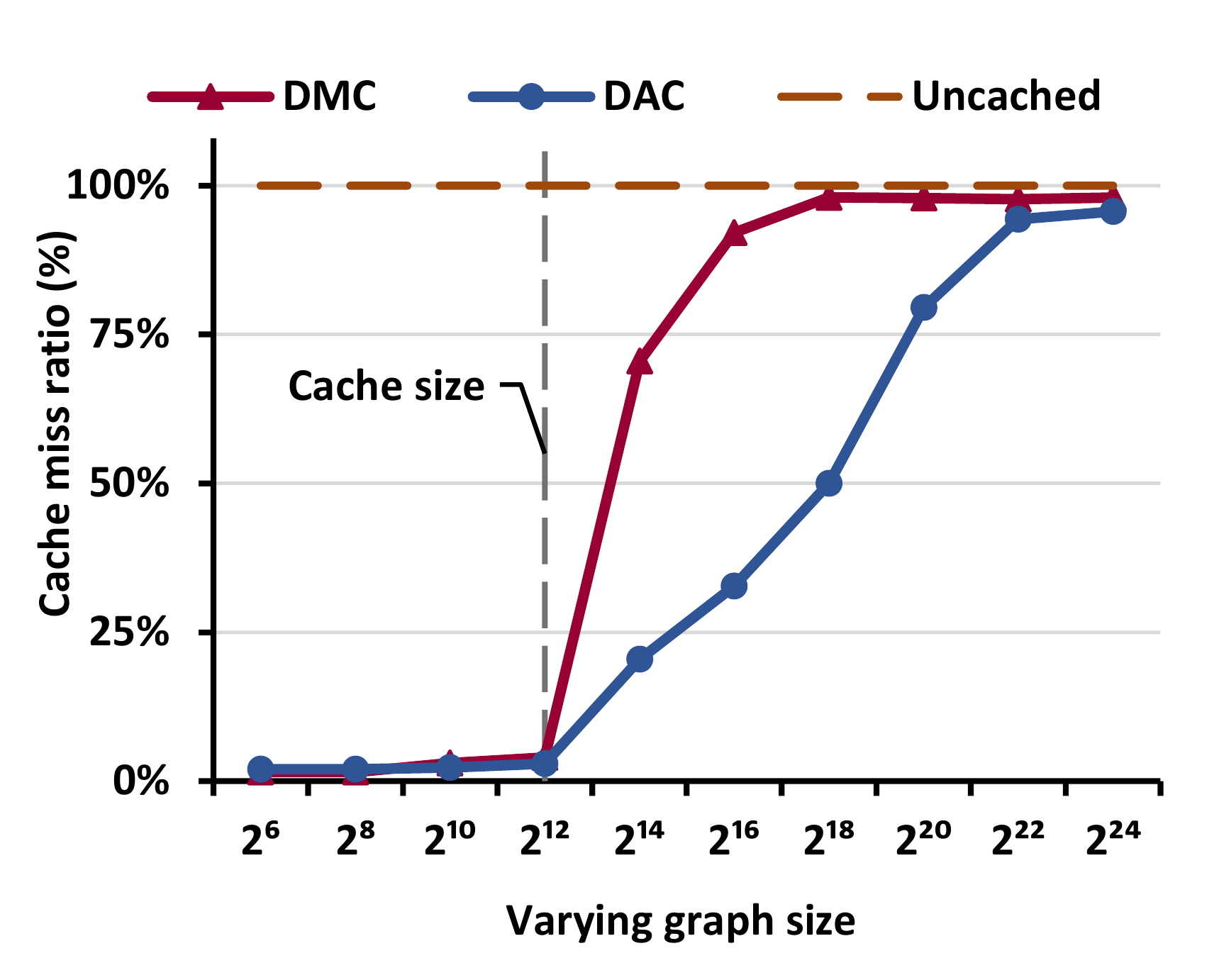}

  \caption{Comparison of the cache miss ratio between DAC and DMC as the size of graphs increases.}
  \label{fig:cache}
\end{minipage}
\end{figure}

\begin{figure*}[t]
\begin{minipage}{.67\linewidth}
 \includegraphics[valign=t,width=\linewidth]{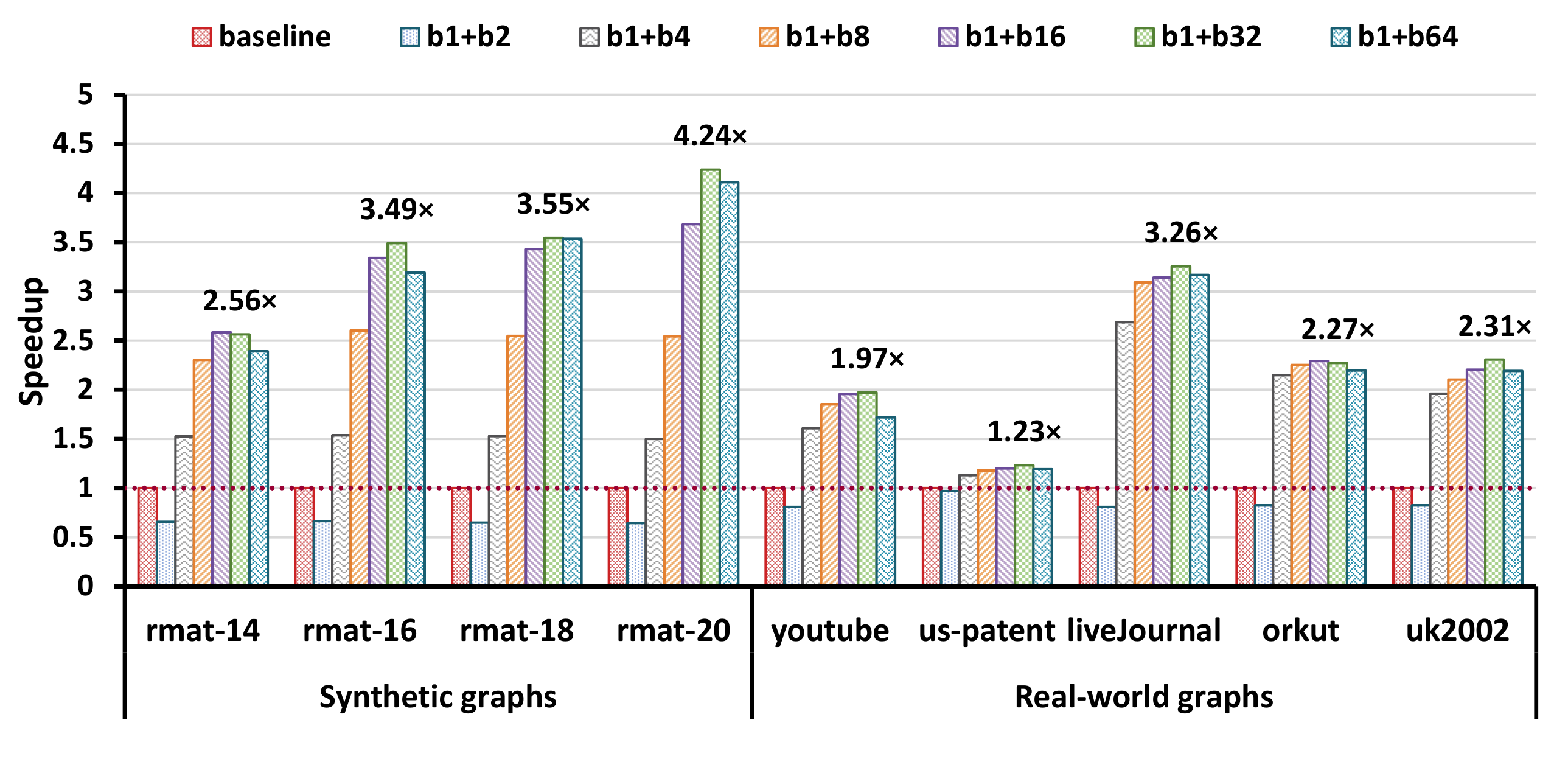}

\vspace{6pt}
  \caption{Speedup of different dynamic burst strategies on MetaPath with synthetic graphs and real world graphs.}
  \label{fig:dyb}

\end{minipage}\hfill
\begin{minipage}{.31\textwidth}
\vspace{12pt}
\subcaptionbox{MetaPath.\label{fig:breakdown_metapath}}
    {\includegraphics[width=\linewidth]{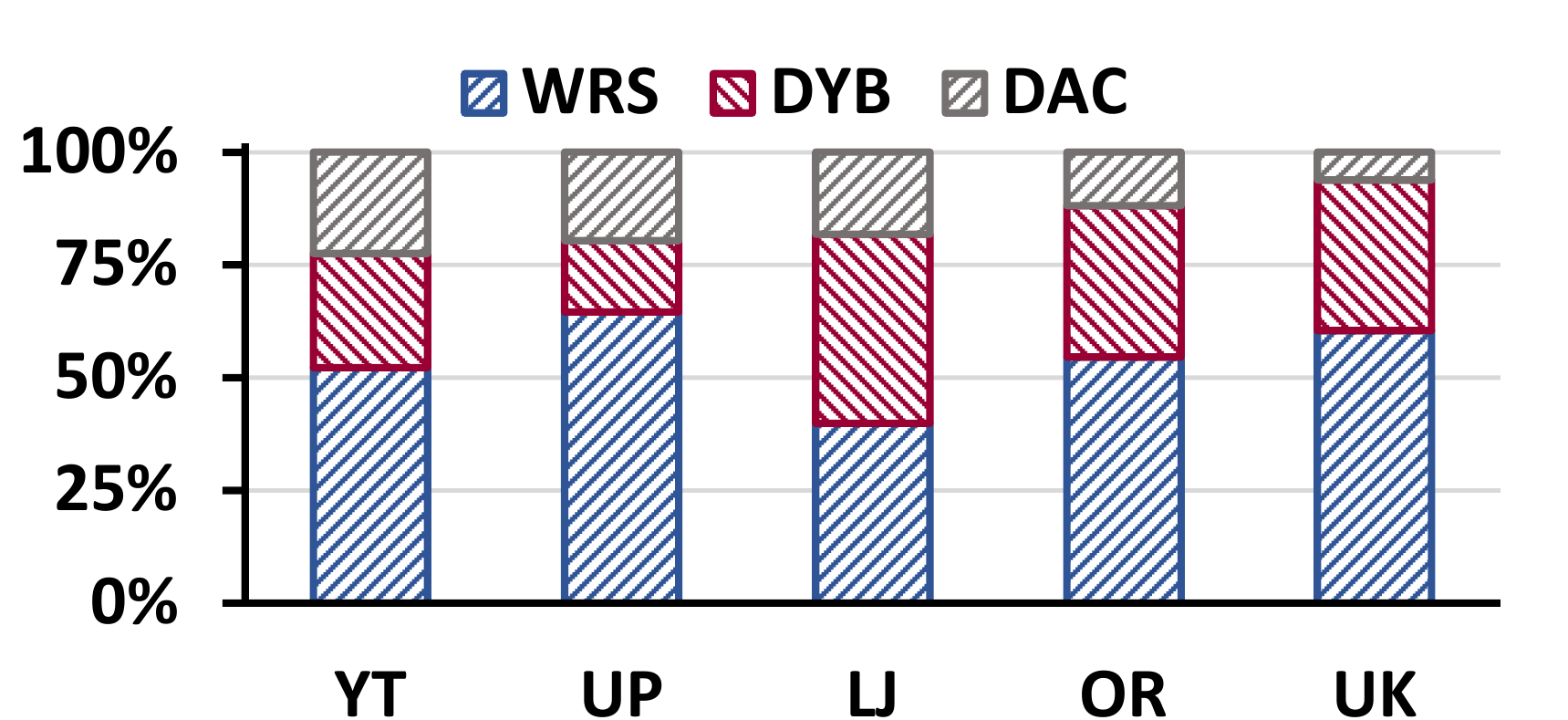}}\vspace{-3pt}
  \subcaptionbox{Node2Vec.\label{fig:breakdown_node2vec}}
    {\vspace{-5pt}\includegraphics[width=\linewidth]{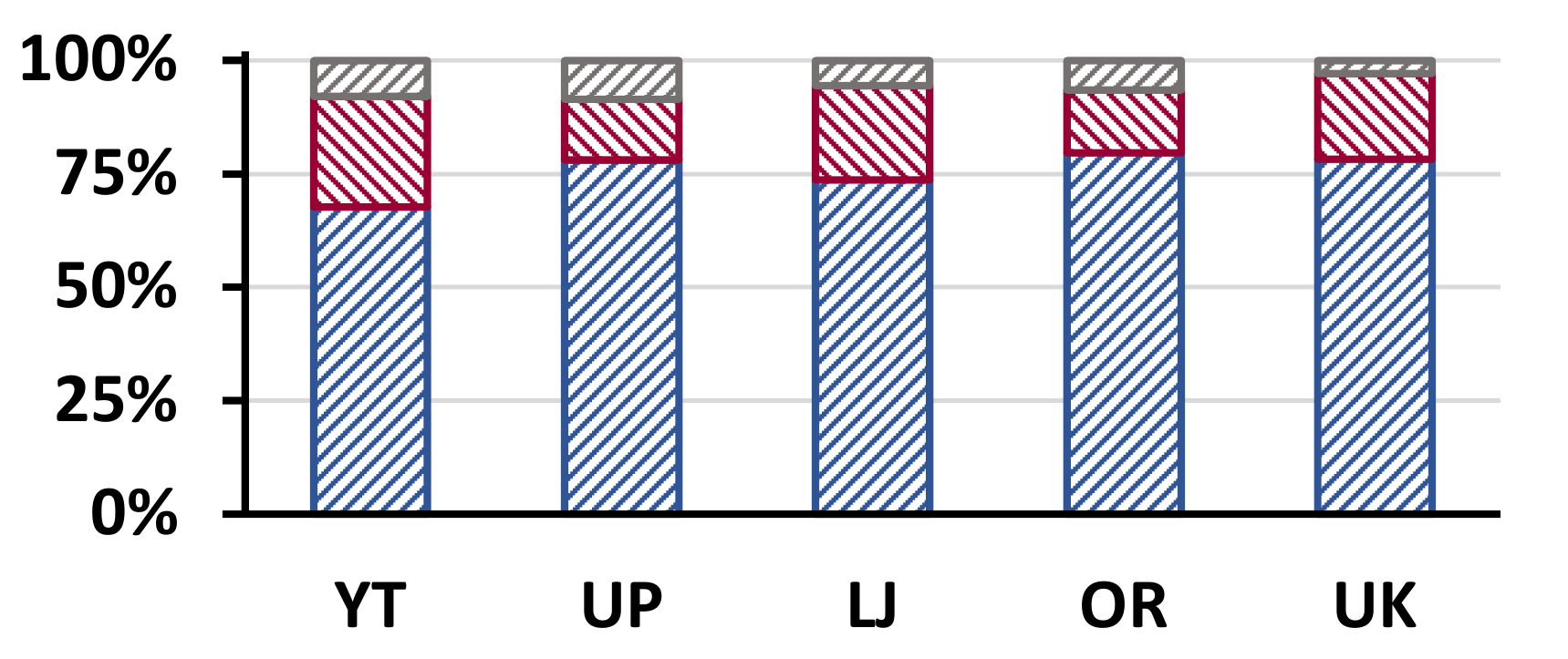}}

\caption{Performance breakdown of WRS, DYB, and DAC on MetaPath and Node2Vec.}
\label{fig:breakdown}
\end{minipage}
\end{figure*}

\setword{The}{rv:r2o1} throughput of the WRS Sampler determines the overall performance of LightRW. Therefore, we conduct an evaluation of the WRS Sampler module with different degrees of parallelism and varying workloads individually. We randomly generate weights and items as the input data to be sampled by the WRS Sampler. The pre-generated data is stored in the one FPGA's DRAM. During the evaluation, the weights are sent to the WRS Sampler in a stream, and we measure the number of traversed items per second as throughput.

\Cref{fig:pwrs_k} shows the throughput of the proposed WRS Sampler with varying degrees of parallelism, $k$ (the number of vertices consumed per cycle). The blue line shows the measured throughput, while the gray dashed line shows the theoretical throughput. We observe that the throughput of the WRS Sampler increases linearly with $k$ and matches the theoretical throughput when $k\leq 16$. At $k=16$, the sampling throughput reaches the maximum bandwidth of the FPGA DRAM ($17$ GB/s), indicating that the proposed sampler can fully utilize the available memory bandwidth. \setword{Moreover, the results demonstrate}{rv:r2o5} the good scalability of our proposed parallel weighted reservoir algorithm and hardware architecture design.

\Cref{fig:pwrs_size} shows the throughput of the WRS Sampler with varying workloads.
Specifically, we generate items of different sizes to be sampled in data streams ranging from $2^6$ to $2^{16}$. To ensure the full bandwidth utilization of the FPGA's DRAM by the WRS Sampler, we set the degree of parallelism to $16$.
The results show that the throughput of the WRS Sampler is slightly less than the maximum memory throughput when the workload size is small. This is due to the initialization overhead of the pipeline execution, but such a performance difference is negligible. Overall, the proposed WRS Sampler is capable of fully utilizing the memory bandwidth with varying workloads.
In other words, the WRS Sampler is not the bottleneck of the entire accelerator pipeline.

\subsection{Impact of Memory Optimizations}

\label{sub:memory_opt}
In this section, we first evaluate the impact of different dynamic burst strategies in the dynamic burst engine. 
We then demonstrate the benefits of our degree-aware cache in reducing the overhead of random accesses in GDRWs.

\subsubsection{Degree-aware Cache}
\label{sub:dac_performance}

\Cref{fig:cache} depicts the cache miss ratio of the proposed degree-aware cache (DAC) compared to the direct-mapped cache (DMC) for MetaPath on RMAT graphs with an increasing number of vertices from $2^6$ to $2^{24}$. The evaluated caches are capable of storing up to $2^{12}$ vertices.

We can make the following observations.
First, for graphs with fewer than $2^{12}$ vertices, the cache miss ratio is close to zero, as all vertices can be cached.
Second, our degree-aware cache has a much lower cache miss ratio than a direct-mapped cache. 
Specifically, as the size of the graph increases, the cache miss ratio of the direct-mapped cache grows significantly, while our cache still maintains a comparatively lower miss ratio.
For example, when the graph has $2^{18}$ vertices, the cache miss ratio of the direct-mapped cache is close to $100\%$, while our degree-aware cache only has a $49\%$ cache miss.

\subsubsection{Dynamic Burst Engine}
\label{sub:dybee}

\Cref{fig:dyb} shows the throughput comparison on MetaPath with different dynamic burst strategies.
The burst strategies are represented as `\text{b\{x\} + b\{y\}}'.
For example, given a strategy \text{b1 + b16}, it indicates that the length of a short burst and a long burst is set to 1 and 16, respectively.
The baseline is \text{b1 + b0}, i.e., the burst length of the access pipeline is only configured to one, which is the common setting to handle irregular workloads~\cite{8892073}.
We set the length of the short burst to one since it achieves minimal overhead in loading unused data, as we have discussed in~\Cref{sub:dynamic_burst_engine}.
Based on the results in ~\Cref{fig:dyb}, we have the following highlights.
First, performance varies significantly with different strategies. The burst strategy \text{b1 + b32} achieves the best performance and outperforms the baseline up to $4.24 {\times}$.
Second, the strategy \text{b1 + b2} delivers the worst performance, because the benefits of long burst with length of two cannot amortize the overhead of the burst plan generation.
Third, although similar trends are observed on both synthetic and real-world graphs, the performance improvements from dynamic burst strategies on real-world graphs are not as significant. This is because real-world graphs are generally more irregular than synthetic graphs. Nonetheless, our dynamic burst strategies can still provide up to $3.26{\times}$ throughput improvement compared to the baseline solution.
As the strategy \text{b1 + b32} outperforms others in all graphs, we use it as the dynamic burst strategy for the subsequent evaluation.

\subsection{Performance Breakdown}\label{sub:performance_breakdown}

\Cref{fig:breakdown} analyzes the performance breakdown of the three proposed techniques, weighted reservoir sampling (WRS), dynamic burst engine (DYB), and degree-aware cache (DAC). Specifically, each proposed technique is disabled one at a time, and the resulting impact on performance is measured relative to the implementation with all techniques enabled.
On the basis of the results, we have the following observations.
First, WRS, which enables pipelined execution, contributes the most performance improvement (up to $41\%$ to $79\%$) for the two GDRW algorithms, particularly for Node2Vec.
Second, the benefit from the dynamic burst engine on Node2Vec is comparatively small. This is because Node2Vec requires additional memory access on the neighbors of the last traversed vertex, which in turn decreases the available memory bandwidth for the dynamic burst engine.
Third, the degree-aware cache contributes more improvement to MetaPath than to Node2Vec. This is because that the degree-aware cache aims to improve the performance on accessing the \emph{col\_index} array, but Node2Vec has more memory accesses on the \emph{row\_index} array.
Even though our cache only occupies several URAMs, it still achieves up to $6\%$ improvement on the largest graph \emph{uk2002}, demonstrating the effectiveness of the proposed cache replacement policy.

\subsection{Comparison with the State-of-the-art CPU Implementation} %
\label{sub:comparison_with_state_of_the_art_cpu_imp}

We compare LightRW with the state-of-the-art CPU-based implementation, ThunderRW~\cite{sun2021thunderrw}. ThunderRW is executed on a server equipped with an Intel Xeon Gold 6246R CPU, which has 16 physical cores and a 19-stage pipeline, with a total cache capacity of 35.75MB~\cite{arafa2019cascade}.\setword{}{rv:r3o1}

First, we present the speedup of MetaPath and Node2Vec on five real-world graphs.
Next, we compare the latency of executing small number of queries.
Finally, we conduct a sensitivity evaluation by varying the query length and number of queries.

\subsubsection{Throughput comparison.}
\label{ssub:throughputcomparsion}

\begin{figure}[t]
\centering
\begin{subfigure}[t]{0.49\linewidth}
\centering
\includegraphics[width=\linewidth]{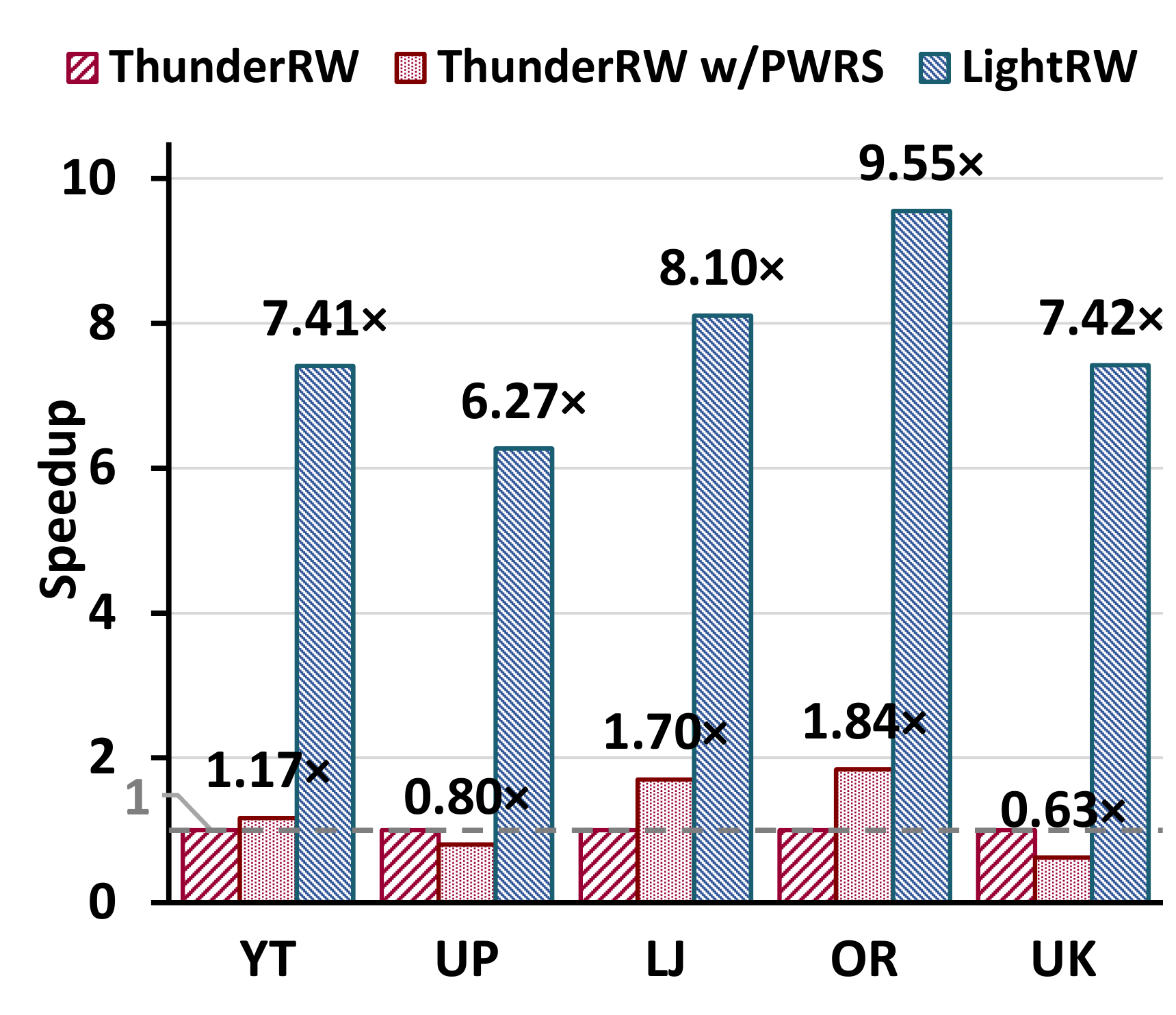}
\vspace{-5pt}
\caption{MetaPath.}
\label{fig:res_meta}
\end{subfigure}
\begin{subfigure}[t]{0.49\linewidth}
\centering
\includegraphics[width=\linewidth]{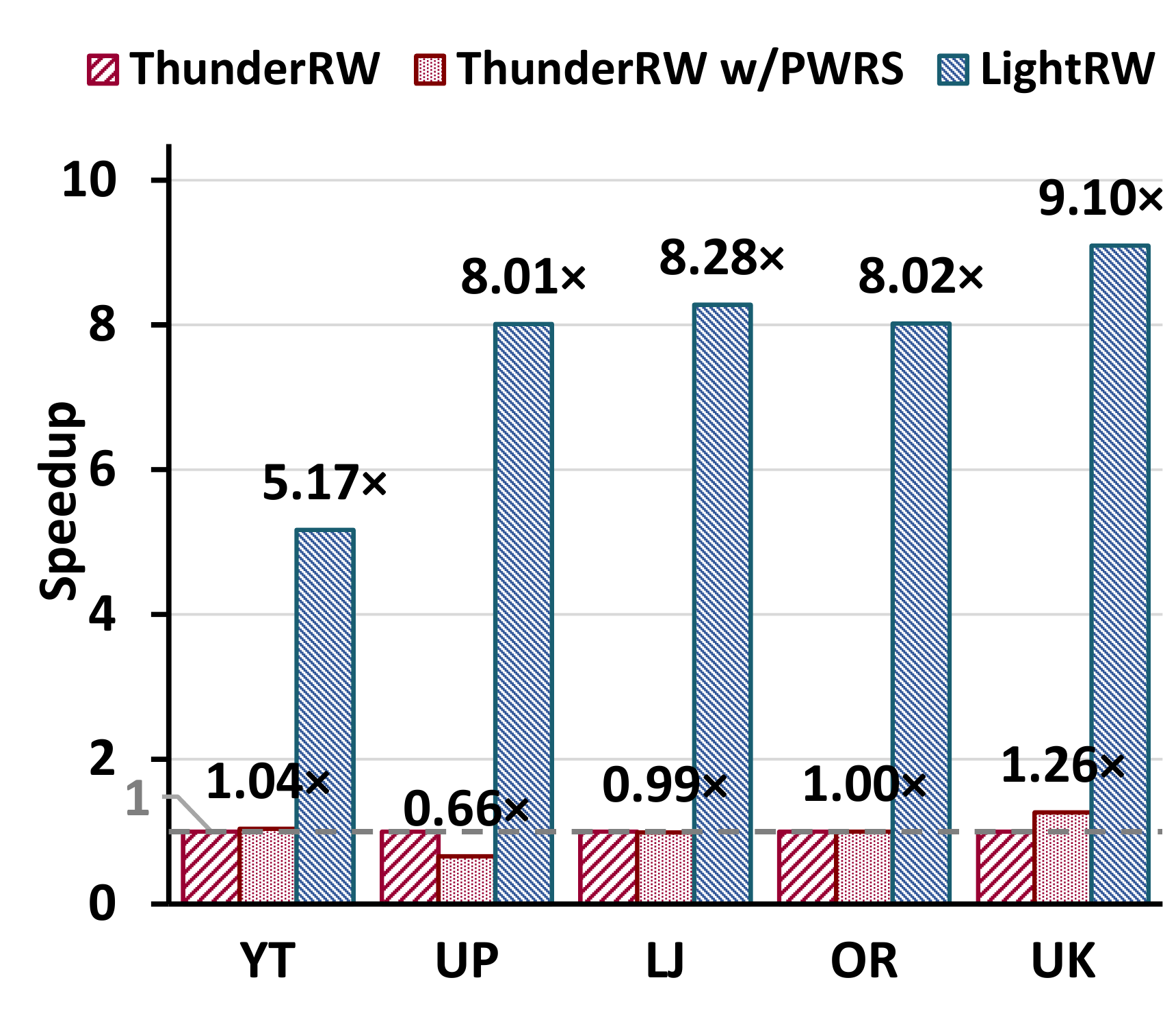}
\vspace{-5pt}
\caption{Node2Vec.}
\label{fig:res_node2vec}
\end{subfigure}
\caption{\rv{Performance comparison between LightRW and ThunderRW on MetaPath and Node2Vec.}}
\label{fig:throughputcomparsion}
\vspace{-3pt}
\end{figure}
\begin{figure}[t]
\centering
\begin{subfigure}[t]{0.49\linewidth}
\centering
\includegraphics[width=\linewidth]{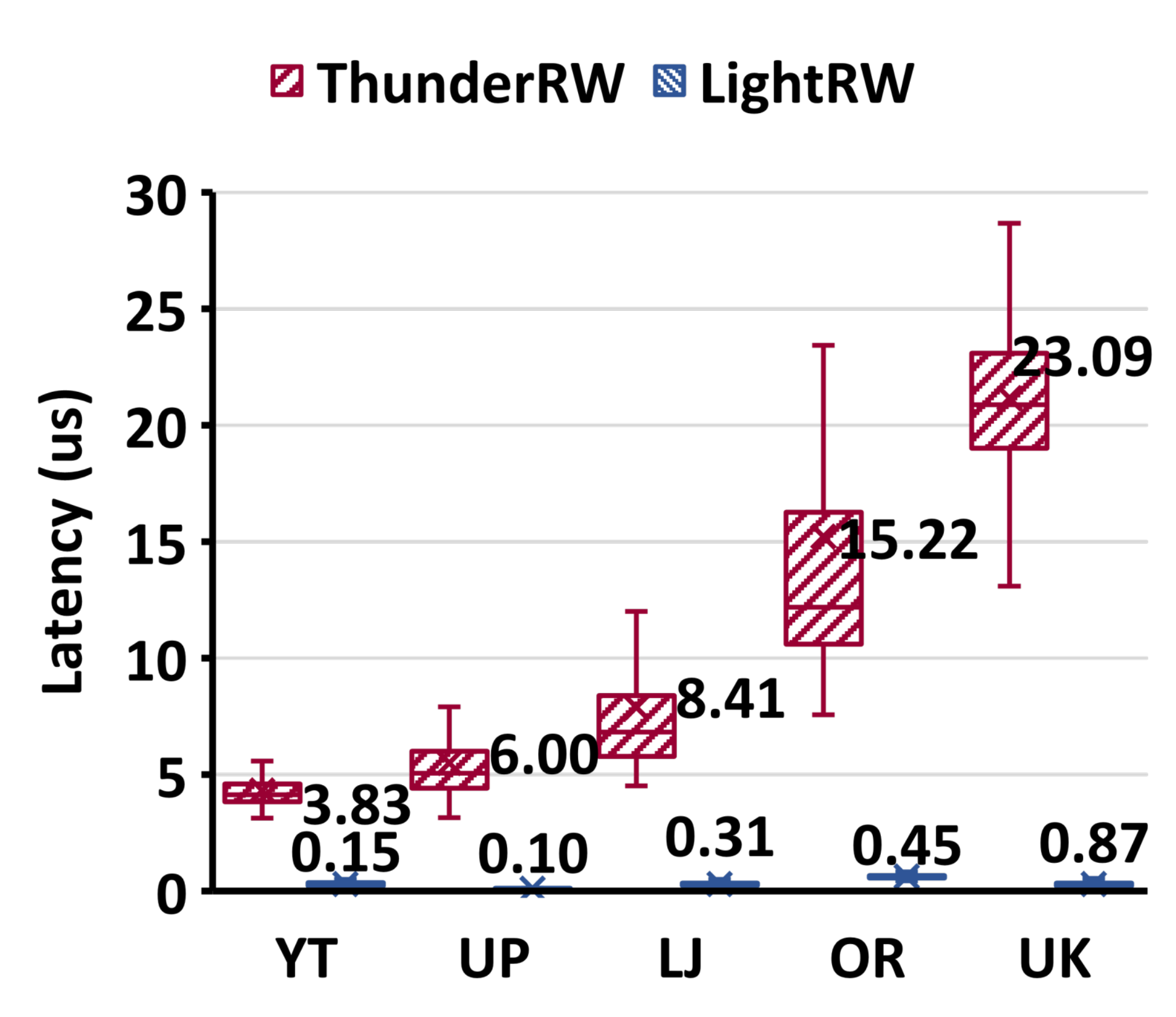}
\vspace{-5pt}
\caption{MetaPath.}
\label{fig:latency_metapath}
\end{subfigure}
\begin{subfigure}[t]{0.49\linewidth}
\centering
\includegraphics[width=\linewidth]{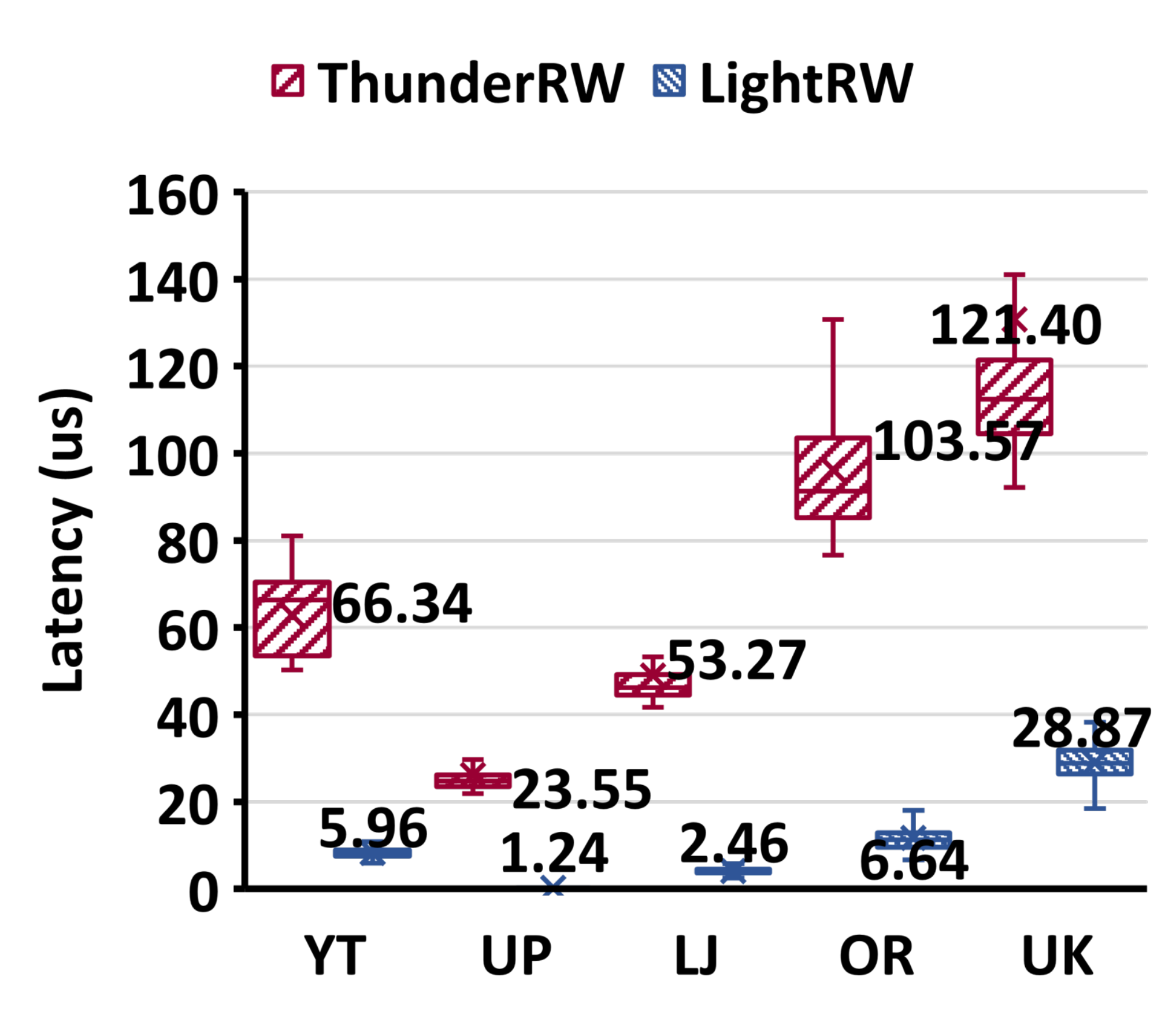}
\vspace{-5pt}
\caption{Node2Vec.}
\label{fig:latency_node2vec}
\end{subfigure}
\caption{Comparison of latency between LightRW and ThunderRW on MetaPath and Node2Vec.}
\label{fig:latency}
\vspace{-3pt}
\end{figure}

\Cref{fig:throughputcomparsion} presents the speedup in performance of LightRW over ThunderRW for both MetaPath and Node2Vec. \rv{Besides the explicit memory prefetching, task level parallelism, and parallelized PRNG~\cite{saito2008simd} that have already been enabled in ThunderRW, we further implement and integrate the proposed parallel weighted reservoir sampling algorithm into ThunderRW (ThunderRW w/PRWS) to make a fair comparison with LightRW.}

\rv{The impact of parallel WRS algorithm on ThunderRW varies on different graphs. For example, $1.84\times$ speedup is observed on \emph{OR} dataset on MetaPath random walk, while significant performance degradation is also observed on the \emph{UP} and \emph{UK} datasets.
There are two potential reasons. First, the maximum number of CPU pipeline stages is limited to 19, which is not suitable for the proposed parallel WRS algorithm that relies on long pipeline execution.
Second, the computational cost of random number generation is not sufficiently amortized over the benefit of reduced memory accesses.}
LightRW outperforms ThunderRW, even running at a $10{\times}$ slower frequency and $2{\times}$ lower sequential memory bandwidth.
In particular, LightRW delivers $6.27{{\times}}\unsim9.55{\times}$ speedup over ThunderRW on MetaPath and $5.17{\times}\unsim9.10{\times}$ speedup on Node2Vec.
This is attributed to the proposed fine-grained pipelining with WRS and memory optimizations.
It is worth noting that the speedup on the \emph{youtube} graph is smaller than others because it is small enough to fit into the CPU last-level cache.

\subsubsection{Latency evaluation}

\begin{figure}[t]
\centering
\begin{subfigure}[t]{0.49\linewidth}
\centering
\includegraphics[width=\linewidth]{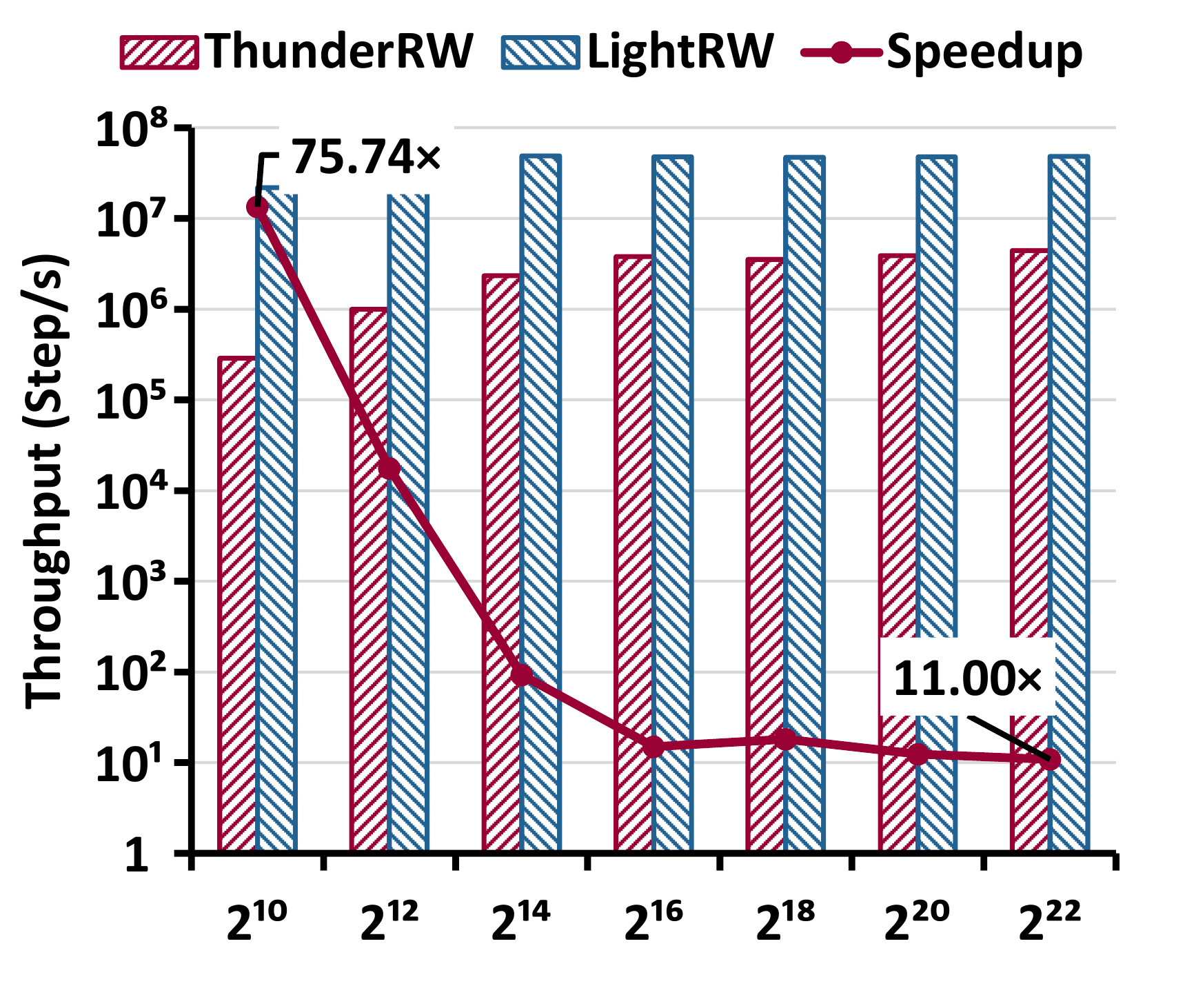}
\vspace{-5pt}
\caption{MetaPath.}
\label{fig:number_query_metapath}
\end{subfigure}
\begin{subfigure}[t]{0.49\linewidth}
\centering
\includegraphics[width=\linewidth]{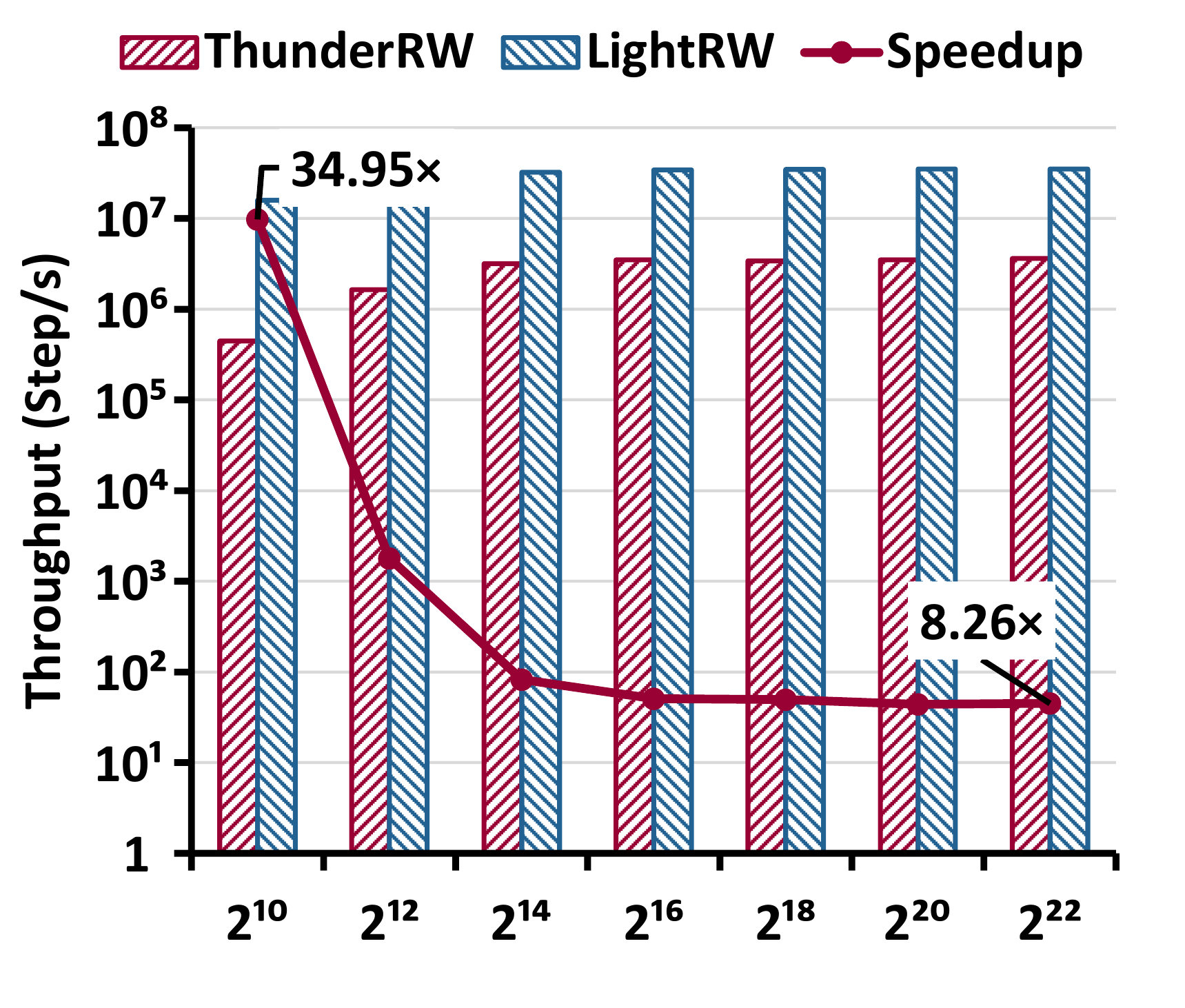}
\vspace{-5pt}
\caption{Node2Vec.}
\label{fig:number_query_node2vec}
\end{subfigure}
\caption{Comparison of throughput between ThunderRW and LightRW on \emph{LJ} with varying numbers of queries on MetaPath and Node2Vec.}
\label{fig:sensitivity_query_num}
\vspace{-3pt}
\end{figure}
\begin{figure}[t]
\centering
\begin{subfigure}[t]{0.49\linewidth}
\centering
\includegraphics[width=\linewidth]{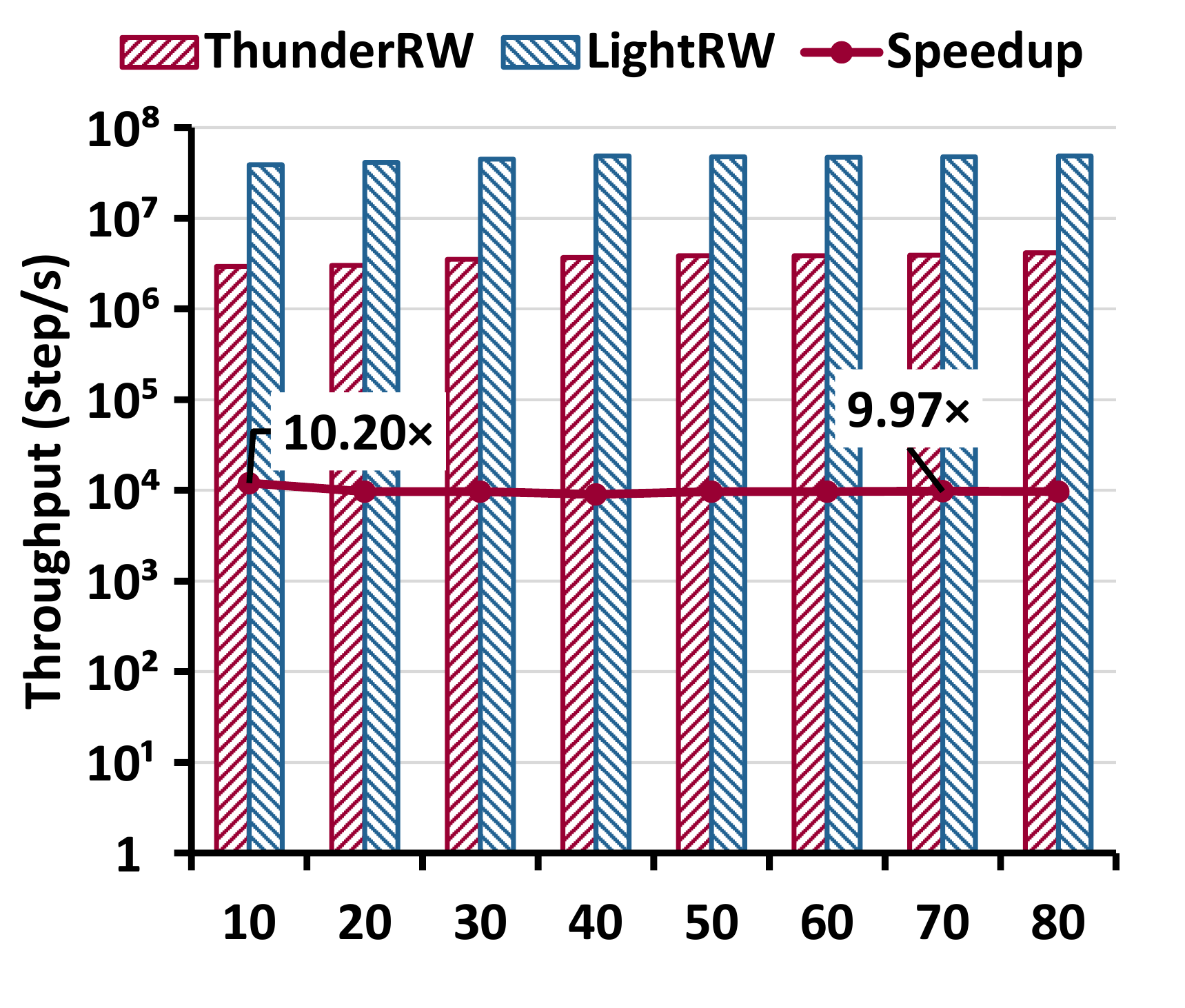}
\vspace{-5pt}
\caption{MetaPath.}
\label{fig:query_length_metapath}
\end{subfigure}
\begin{subfigure}[t]{0.49\linewidth}
\centering
\includegraphics[width=\linewidth]{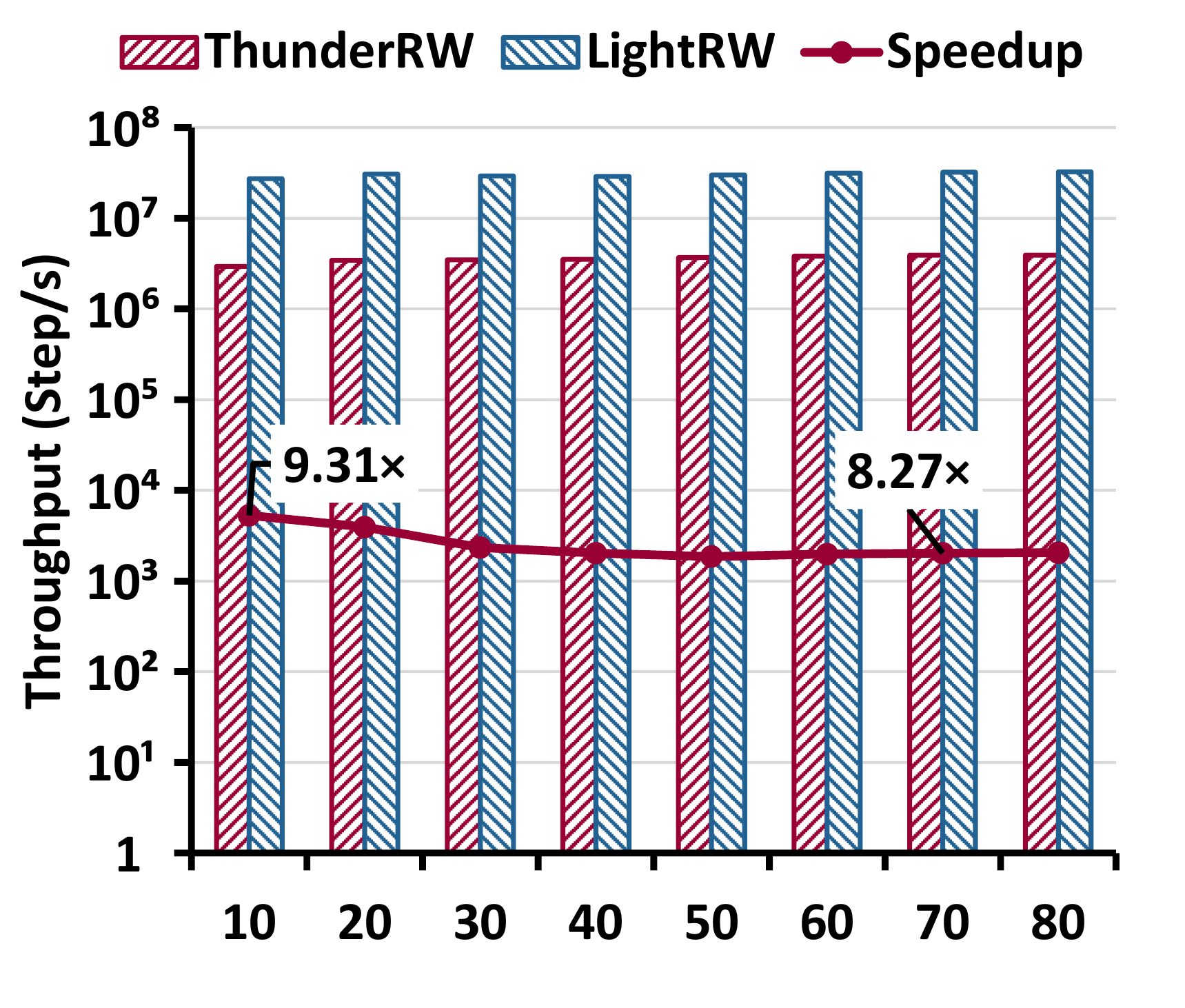}
\vspace{-5pt}
\caption{Node2Vec.}
\label{fig:query_length_node2vec}
\end{subfigure}
\caption{Comparison of throughput between ThunderRW and LightRW on \emph{LJ} with varying query lengths on MetaPath and Node2Vec.}
\label{fig:sensitivity_query_length}
\vspace{-4pt}
\end{figure}

~\Cref{fig:latency} shows the execution latency of LightRW and ThunderRW. Latency is calculated as the time from the start to the end of a query. For LightRW, we implement a cycle counter inside the accelerator to measure the exact hardware cycles. We measure the latency of 8192 randomly selected queries for each graph using both systems. In Figure \ref{fig:latency}, the rectangular box of each data series indicates the range from the lower quartile (25th percentile) to the upper quartile (75th percentile), and the horizontal line inside the box represents the median measurement. The whiskers with the horizontal lines at the top and bottom represent the maximum and minimum measurements, respectively.

On the whole, LightRW achieves much lower latency than ThunderRW.
This suggests that FPGA-based GDRW is more suitable for handling real-time graph analytic applications~\cite{liu2019real,kumar2020graphone,fairbanks2013statistical}.
Additionally, the latency of LightRW is more consistent across different graphs compared to ThunderRW.
This is because the FPGA-based design has deterministic latency, CPU-based designs often suffer from uncontrolled latency and non-deterministic resource sharing, such as hardware interrupt preemption, user/kernel space context switching, and synchronization between multiple threads, etc.

\subsubsection{Sensitivity evaluation on query parameters}
\label{ssub:sensitivity_evaluation_on_query_parameters}

To further investigate the performance of LightRW and ThunderRW, we conduct a sensitivity analysis on a representative dataset, the liveJournal (\emph{LJ}) graph.
This graph has a moderate number of edges and its average degree is the median of all evaluated real-world graphs. Specifically, we vary the number and length of queries in our analysis.

\Cref{fig:sensitivity_query_num} presents the throughput of LightRW and ThunderRW with the number of queries ranging from $2^{10}$ to $2^{22}$ of random walk on the \emph{LJ} graph. We start the number of queries from $2^{10}$ because ThunderRW requires at least $2^{10}$ queries to fully utilize the memory bandwidth of the target CPU.
First, we can see that LightRW delivers almost constant throughput regardless of the number of queries. The throughput is up to $4.8{\times}10^{7}$ steps per second on MetaPath and $3.5{\times}10^{7}$ steps per second on Node2Vec.
Next, we observe that LightRW's speedup over ThunderRW is $11.00{\times}\unsim75.74{\times}$ on MetaPath and $8.26{\times}\unsim34.95{\times}$ on Node2Vec.
In particular, the speedup is more significant when the number of queries is relatively small (e.g., up to $75.54\times$ with $2^{10}$ queries).
This is because ThunderRW has constant initialization overhead, such as memory allocation and thread allocation, before the execution of multiple queries.

\Cref{fig:sensitivity_query_length} shows the throughput of LightRW and ThunderRW with varying query lengths from $10$ to $80$.
The trends show that LightRW delivers constant throughput with different query lengths.
The performance speedup of LightRW over ThunderRW is around $9.97{\times}\unsim10.20{\times}$ for MetaPath, and $8.28{\times}\unsim9.31{\times}$ for Node2Vec.
Therefore, we conclude that LightRW stably outperforms ThunderRW on MetaPath and Node2Vec with different query settings.

\subsubsection{Comparison of power efficiency}
\label{ssub:comparison_of_power_efficiency_and_monetary_cost}

\begin{table}[t]
\centering
\caption{The comparison of power efficiency between LightRW and ThunderRW on all tested graphs.}
\label{tab:power}

\begin{threeparttable}
\begin{tabularx}{\linewidth}{lcccc}
\toprule
\multicolumn{1}{c|}{\multirow{2}{*}{{\text{App.}}}}
& \multicolumn{2}{c}{\text{Power consumption (Watt)}}
& \multirow{1}{*}{\text{Power efficiency}}
\\
\multicolumn{1}{c|}{}
& \multirow{1}{*}{{LightRW}}
& \multirow{1}{*}{{ThunderRW}}
&\multirow{1}{*}{\text{improvement}}
\\

\midrule

\multicolumn{1}{c|}{MetaPath} & $41\unsim45$  & $103\unsim124$ & $15.05{\times}\unsim26.42{\times}$ \\
\multicolumn{1}{c|}{Node2Vec} & $39\unsim42$  & $110\unsim126$ & $16.28{\times}\unsim24.10{\times}$  \\

\bottomrule
\end{tabularx}%

\end{threeparttable}

\end{table}

\Cref{tab:power} compares the power consumption of LightRW and ThunderRW. The power consumption of CPU-based execution is measured using the CPU Energy Meter~\cite{cpumeter} during the execution of each benchmark, while the power consumption of our LightRW FPGA accelerator is measured using {xbutil}~\cite{vitis}.

The power efficiency improvement is calculated as the ratio of end-to-end execution time per Watt of LightRW to that of ThunderRW.
Overall, LightRW consumes less energy than ThunderRW and outperforms ThunderRW by $15{\times}\unsim26{\times}$ on MetaPath and $16{\times}\unsim24{\times}$ on Node2Vec in power efficiency.
\rv{\setword{Since}{rv:r2o4} the power of LightRW on the FPGA platform is only $2.29{\times}$ less than that of ThunderRW on CPUs, and LightRW runs to $9.55\times$ fast ThunderRW,  we conclude that the improvement in power efficiency is mainly due to our design rather than a power-efficient hardware platform.}
The significant performance and energy efficiency improvements also demonstrate the efficacy of customizing accelerators for GDRWs.

\subsection{Other Results}

In this section, we analyze the PCIe overhead of FPGA acceleration and present the hardware resource utilizations of LightRW for different GDRW algorithms

\subsubsection{PCIe Overhead Analysis}

\begin{table}[h]
\centering
\caption{The proportion of PCIe data transfer time over the end-to-end execution time of MetaPath and Node2Vec.}
\label{tab:pcie_time}

\begin{threeparttable}
\begin{tabularx}{\linewidth}{lccccc}
\toprule
\multicolumn{1}{c|}{\multirow{1}{*}{\text{App.}}}

& \multirow{1}{*}{youtube}
& \multirow{1}{*}{us-patents}
& \multirow{1}{*}{liveJournal}
& \multirow{1}{*}{orkut}
& \multirow{1}{*}{uk2002}
\\

\midrule
\multicolumn{1}{c|}{\multirow{1}{*}{{MetaPath}}} & $16.5\%$ & $15.3\%$ & $20.5\%$  & $33.5\%$ & $23.3\%$  \\
\multicolumn{1}{c|}{\multirow{1}{*}{{Node2vec}}} & $0.07\%$& $1.10\%$& $0.54\%$& $0.56\%$& $0.25\%$ \\
\bottomrule
\end{tabularx}%

\end{threeparttable}

\end{table}

\Cref{tab:pcie_time} shows the percentage of time taken for PCIe data transfer over the end-to-end execution time.
The results demonstrate that graph data transfer generally occupies only a small portion of the total end-to-end execution time, ranging from $0.07\%$ to $33.5\%$. This suggests that data transfer is not a bottleneck for the entire system, and justifies the offloading of GDRWs to the FPGA accelerator.

\subsubsection{FPGA resource utilization}

\begin{table}[h]
\centering
\caption{The consumption of hardware resources (percentage) and frequency (MHz) of MetaPath and Node2Vec on Alveo U250 FPGA borad.}
\label{tab:resource}

\begin{threeparttable}
\begin{tabularx}{\linewidth}{lXXXXXX}
\toprule
\multicolumn{1}{c|}{\multirow{1}{*}{{\text{App.}}}}
& \multirow{1}{*}{{LUTs}}
& \multirow{1}{*}{{REGs}}
& \multirow{1}{*}{{BRAMs}}
& \multirow{1}{*}{{DSPs}}
& \multirow{1}{*}{{Frequency}}
\\

\midrule

\multicolumn{1}{c|}{MetaPath} & $33.52\%$  & $29.76\%$ & $17.24\%$ & $5.16\%$ &300MHz \\
\multicolumn{1}{c|}{Node2Vec} & $20.84\%$  & $18.20\%$ & $36.12\%$ & $2.62\%$ &300MHz \\

\bottomrule
\end{tabularx}%

\end{threeparttable}

\end{table}

\Cref{tab:resource} presents the resource utilization and frequency of LightRW for MetaPath and Node2Vec. The results show that LightRW utilizes only a small portion of the resources available on the U250 FPGA, leaving ample resources for downstream processing logic, such as graph learning applications. Additionally, LightRW runs at a frequency of up to $300$MHz benefiting from the modular design methodology used.

\subsection{\rv{Case Study: Link Prediction}}
\label{sub:case_study_link_prediction_with_node2vec}

\rv{As a case study, we integrated our accelerator into the Stanford Network Analysis Platform (SNAP) to accelerate the link prediction application~\cite{enwiki:1100501520}. SNAP is a popular graph analysis platform with a core library written in C++, optimized for maximum performance and compact graph representation~\cite{leskovec2016snap}. Link prediction is a widely used function in applications such as social networks, bioinformatics, and e-commerce~\cite{enwiki:1100501520}. The goal is to predict the existence of a link between two entities in a given network. A common approach is to use the Node2Vec random walk to generate different paths, followed by an unsupervised learning method, such as Word2Vec~\cite{grover2016node2vec}, to process the random walk paths and identify vertices with minimal cosine similarity to form the predicted edges.
}

\begin{figure}[t]
     \centering
     \includegraphics[width=\linewidth]{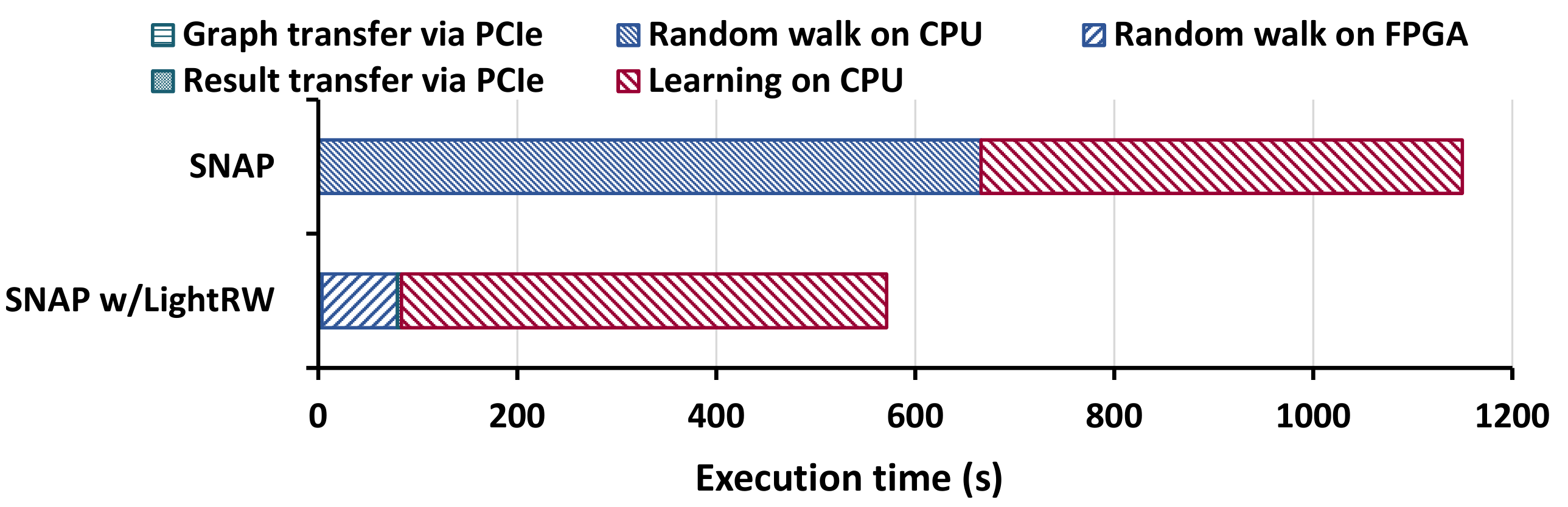}
     \caption{\rv{Execution time breakdown for link prediction on \emph{LJ} graph between SNAP and LightRW accelerated SNAP (SNAP w/LightRW)}}
     \label{fig:cs}
\end{figure}

\rv{\Cref{fig:cs} presents the execution time breakdown of LightRW-accelerated SNAP (SNAP w/LightRW) and CPU-based SNAP for link prediction on the liveJournal graph. The results show that the Node2Vec random walk is the most time-consuming part of the link prediction application. With the acceleration of the Node2Vec random walk by LightRW, the total execution time of the link prediction is reduced to half the original execution time. Furthermore, although graph data and results need to be copied between CPU and FPGA memory, the impact is negligible compared to the overall execution time. The results also indicate the potential for further performance improvements if the learning process can also be offloaded to the FPGA side for acceleration.
}

\section{Related Works}
\label{sec:related_works}

\noindent\textbf{Graph Random Walks on CPU/GPUs.}
Knightking~\cite{yang2019knightking} adopts a vertex-centric model for a single query and processes multiple queries with the BSP execution model, i.e., waiting for all queries to complete the current step before starting the next step. %
To leverage the parallel processing capability of GPUs, C-SAW~\cite{pandey2020c} parallelizes the initialization and generation stage using the inverse transformation sampling method among multiple GPU processing cores. Multiple queries are processed using the BSP model.
ThunderRW~\cite{sun2021thunderrw} proposes a step-centric model that interleaves memory accesses of subtasks to improve the parallelism of multiple queries, achieving state-of-the-art performance and outperforming the GPU-based C-SAW.

\noindent\textbf{Graph Random Walks on FPGAs.}
\citeauthor{su2021graph}~\cite{su2021graph} propose an FPGA-based accelerator for static random walks that utilizes {\em high bandwidth memory} (HBM) and multiple processing elements to handle multiple queries concurrently. They develope a degree-aware sampler that selects the appropriate division arithmetic units based on the degree of the vertices. This work is designed to support a specific MetaPath query by partitioning the graph into subgraphs based on vertex labels and performing a static random walk on each subgraph.
\rv{However, their architecture is highly specific to MetaPath queries with only two types of vertex labels. Because it only supports uniform random sampling, the proposed accelerator is limited to unweighted MetaPath random walks and cannot be generalized to any other GDRW algorithms. On the other hand, LightRW is an accelerator for dynamic random walks and supports not only generic MetaPath queries but also other GDRW algorithms. Our performance improvements are mainly due to the novel FPGA architecture design that efficiently realizes the proposed parallel weighted reservoir sampling algorithms.}

\section{Conclusion and Future Work}
\label{sec:conclusion}
This paper proposes LightRW, the first FPGA-based accelerator for GDRWs. Unlike existing CPU-based approaches that require synchronization barriers among different stages, LightRW parallelizes the weighted reservoir sampling for GDRWs to enable a fine-grained pipeline execution on the chip, allowing for better spatial parallelism and reduced memory access to the DRAM. Additionally, LightRW contains two novel memory optimizations to handle memory access patterns specific to random walks with better efficiency.
Experimental results demonstrate that LightRW achieves up to $9.55{\times}$ and $9.10{\times}$ performance speedup compared to the state-of-the-art CPU-based implementation on two representative dynamic random walk algorithms. However, processing large graphs (e.g., in Terabyte scale) may require multiple FPGA boards with sufficient computation power and DRAM.

\emph{Future work:} First, we plan to develop a distributed version of LightRW to leverage the increased availability of high-speed network interfaces (e.g., InfiniBand and 100G Ethernet) and open-source network frameworks on FPGAs (e.g., OpenNIC~\cite{opennic} and Corundum~\cite{forencich2020fccm}). Second, \rv{
\setword{LightRW}{rv:r1o4} demonstrates the advantage of a fine-grained pipelined accelerator over generic CPU hardware. Specifically, our pipelined architecture eliminates the high cost of the random number generation stage and the weighted random sampling initialization stage. This idea is not limited to GDRWs but further enables the possibility of more efficient computing for both graph-based and non-graph-based applications that rely on randomized approaches, such as Markov chain Monte Carlo~\cite{Banerjee_ASPLOS2019}, Bayesian networks~\cite{dellaportas2002bayesian}, and physics simulations~\cite{plimpton2019direct}.
}

\begin{acks}
{
This research/project is supported by the National Research Foundation, Singapore under its AI Singapore Programme (AISG Award No: AISG2-TC-2021-002), the Ministry of Education AcRF Tier 2 grant (No. MOE-000242-00 / MOE-000242-01),
and Google South \& Southeast Asia Research Award 2022. We also thank the AMD Heterogeneous Accelerated Compute Clusters (HACC) program (formerly known as the XACC program - Xilinx Adaptive Compute Cluster program)~\cite{hacc} for the generous hardware donation.
Bingsheng He and Yao Chen from the National University of Singapore are the corresponding authors.
}
\end{acks}

\bibliographystyle{ACM-Reference-Format}
\bibliography{reference}

\end{document}